\documentclass[twocolumn]{aastex61}

\received{}
\revised{}
\accepted{}
\submitjournal{ApJ}

\shorttitle{SMA Observations of NGC 3627}
\shortauthors{Law et al.}

\begin{document}

\title{SMA Observations of Extended $\rm{CO}\,(J=2-1)$ Emission in Interacting Galaxy NGC 3627}

\correspondingauthor{Charles J. Law}
\email{charles.law@cfa.harvard.edu}

\author{Charles J. Law}
\affiliation{Harvard-Smithsonian Center for Astrophysics, 60 Garden St., Cambridge, MA 02138, USA}

\author{Qizhou Zhang}
\affiliation{Harvard-Smithsonian Center for Astrophysics, 60 Garden St., Cambridge, MA 02138, USA}

\author{Luca Ricci}
\affiliation{Department of Physics and Astronomy, California State University, Northridge, 18111 Nordhoff Street, 91130, Northridge, CA, USA}

\author{Glen Petitpas}
\affiliation{Harvard-Smithsonian Center for Astrophysics, 60 Garden St., Cambridge, MA 02138, USA}

\author{Maria J. Jim\'{e}nez-Donaire}
\affiliation{Harvard-Smithsonian Center for Astrophysics, 60 Garden St., Cambridge, MA 02138, USA}

\author{Junko Ueda}
\affiliation{Harvard-Smithsonian Center for Astrophysics, 60 Garden St., Cambridge, MA 02138, USA}

\author{Xing Lu}
\affiliation{National Astronomical Observatory of Japan, 2-21-1 Osawa, Mitaka, Tokyo, 181-8588, Japan}

\author{Michael M. Dunham}
\affiliation{Department of Physics, State University of New York at Fredonia, Fredonia, NY 14063, USA}



\begin{abstract}
We present moderate (${\sim}5^{\prime\prime}$) and high angular resolution (${\sim}1^{\prime\prime}$) observations of $^{12}\rm{CO\,}(J=2-1)$ emission toward nearby, interacting galaxy NGC~3627 taken with the Submillimeter Array (SMA). These SMA mosaic maps of NGC~3627 reveal a prominent nuclear peak, inter-arm regions, and diffuse, extended emission in the spiral arms. A velocity gradient of ${\sim}400$--$450$~km~s$^{-1}$ is seen across the entire galaxy with velocity dispersions ranging from $\lesssim 80$~km~s$^{-1}$ toward the nuclear region to $\lesssim 15$~km~s$^{-1}$ in the spiral arms. We also detect unresolved $^{13}\rm{CO\,}(J=2-1)$ line emission toward the nuclear region, southern bar end, and in a relatively isolated clump in the southern portion of the galaxy, while no $\rm{C}^{18}O(J=2-1)$ line emission is detected at a $3\sigma$~rms noise level of 42~mJy~beam$^{-1}$ per 20~km~s$^{-1}$ channel. Using RADEX modeling with a large velocity gradient approximation, we derive kinetic temperatures ranging from ${\sim}5$--$10$~K (in the spiral arms) to ${\sim}25$~K (at the center) and H$_2$ number densities from ${\sim}$400--1000~cm$^{-3}$ (in the spiral arms) to ${\sim}$12500~cm$^{-3}$ (at the center). From this density modeling, we find a total H$_2$ mass of $9.6\times10^9~M_{\odot}$, which is ${\sim}50\%$ higher than previous estimates made using a constant H$_2$--CO conversion factor but is largely dependent on the assumed vertical distribution of the CO gas. With the exception of the nuclear region, we also identify a tentative correlation between star formation efficiency and kinetic temperature. We derive a galactic rotation curve, finding a peak velocity of ${\sim}207$~km~s$^{-1}$ and estimate a total dynamical mass of $4.94 \pm 0.70 \times 10^{10} M_{\odot}$ at a galactocentric radius of ${\sim}6.2$~kpc ($121^{\prime\prime}$).
\end{abstract}




\section{Introduction} \label{sec:intro}

In addition to being fascinating in their own right, nearby galaxies are crucial to the understanding of galaxy evolution and interactions. They allow us to directly resolve the region around active nuclei and individual star-forming clouds over the full galactic disk. This comprehensive view allows us to connect the small-scale physics of the interstellar medium (ISM) and star formation to the disk-wide processes that drive galaxy evolution. It is also well known that interactions and mergers are an important process in galaxy evolution as shown by the increased merger rate in the early Universe \citep[e.g.,][]{Bridge10}. Gravitational interactions between galaxies significantly alter the morphology, luminosity, color, size, star formation rate, and mass distribution in a relatively short period of time. The NGC~3627 system is the natural choice for a representative nearby galaxy and interaction study because of its environmental richness and large suite of complementary data sets ranging from X-ray through radio wavelengths. 

Most of the previous extragalactic molecular gas studies used the $^{12}$CO($J=1-0$) line to trace molecular gas mass. With the recent abundance of $^{12}$CO($J=2-1$) observations, a quantitative understanding of the $^{12}$CO($J=2-1$)  / $^{12}$CO($J=1-0$) line ratio (R$_{21/10}$) is required to compare results across the literature. Adding to the urgency of this, $^{12}$CO($J=2-1$) and $J=3-2$ lines are now regularly observed from z $\sim$ 1--3 galaxies \citep[e.g.,][]{Tacconi13, Tacconi17}, where they are used (with assumptions) to trace the total gas supply. A quantitative understanding of the variation in the R$_{21/10}$ line ratio is required to discern what is driving the changes in R$_{21/10}$ observed on large (galaxy) scales as well as to make rigorous statements about the behavior of molecular gas across galaxy populations.

Additionally, observations of the full $^{12}$CO($J=2-1$) distribution in the very outer arms and inter-arm regions of a spiral galaxy such as NGC~3627 are particularly informative. Typical spiral galaxies do not contain conspicuous $^{12}$CO($J=2-1$) emission in the arm and inter-arm regions as observed in NGC~3627. Comparing this gas tracer with the large existing set of complementary data for NGC~3627 will provide a unique opportunity to study gas conditions in a very wide range of environments.
  
NGC 3627\footnote{A summary of basic astronomical information is presented in Table \ref{tab:basic_info}.} (M66) is a spiral galaxy [RC3 type SAB(s)b; \citealt{Vaucouleurs91}] in the Leo Triplet galaxy group and displays signatures of a LINER/Seyfert 2-type nuclear activity in its spectrum \citep{Ho97, Peng98}. Optical broadband images of NGC~3627 reveal a weak optical bar, two prominent, asymmetric spiral arms, and large-scale dust lanes \citep{Arp66, Ptak06}. The perturbed morphology of the western arm, which appears to be displaced from the plane of the galaxy, provides evidence for recent interaction with neighboring galaxy NGC~3628 \citep[e.g.,][]{Rots78, Haynes79, Soida01}. The close proximity (${\sim}11\rm{\,Mpc}$) and high inclination (${\sim}61^{\circ}$) of NGC~3627 allow for an excellent view onto its spiral structure and pronounced dust patterns, making the galaxy an attractive candidate for investigating post-interaction galactic evolution. As a result, NGC 3627 has been studied in a wide range of continuum and spectroscopic observations -- in \ion{H}{1} \citep{Zhang93, Haan08}, CO (e.g., \citealt{Reuter96, Regan01, Helfer03, Kuno07, Leroy09, Warren10, Morokuma15, Beuther17, Donaire17, Cormier18}), H$\alpha$ \citep{Chemin03}, HCN / HCO$^+$ (\citealt{Krips08, Murphy15, 2017MNRAS.466...49J, Gallagher18}), UV \citep{Calzetti15}, and X-ray emission \citep{Georgantopoulos02, Soida12}. 

Multiple CO and radio continuum (327 MHz, 1.4 GHz, and 2.64 GHz) mapping observations \citep[e.g.,][]{Paladino08, Paladino09, Haan09, Nikiel13} have revealed that the majority of molecular gas in the galaxy is localized in a narrow bar structure ${\sim}300\rm{\,pc}$ in width with emission peaking at the nuclear position and extending along the leading edges of the bar, forming two broad peaks at the bulge ends and trailing off into the spiral arms. However, in atomic \ion{H}{1} emission, NGC 3627 appears to have a spiral structure free of any bar-like signatures \citep{Haan08, Walter08}. An inner ring (${\sim}30^{\prime \prime}$--$60^{\prime \prime}$) is also reported in $^{12}$CO($J=1-0$) \citep{Regan02} and H\textsc{$\alpha$} \citep{Chemin03} observations. An elongated inner ring along the north-south direction is also seen in GALEX far-ultraviolet (FUV) emission \citep[$\lambda_{\rm{eff}}=1516$ \AA;][]{Gil07} and surrounds a net depression in nuclear FUV emission with the $^{12}$CO bar-like structure being contained inside this FUV hole \citep{Casasola11}. Molecular transitions from HCN and HCO$^+$ have also been detected, indicating the presence of high density gas \citep{Gao04, Krips08, Murphy15}. Based on X-ray observations, \citet{Soida12} proposed a recent collision of NGC 3627 and a dwarf companion galaxy to explain some of the spiral arm distortions.

NGC 3627 has a relatively high molecular gas fraction relative to other local star-forming galaxies \citep[e.g.,][]{Casasola04, Saintonge11} with molecular gas mass being comparable to atomic gas \citep{Helfer03, Walter08}. \citet{Zhang93} have suggested that the high H$_2$/\ion{H}{1} mass ratio of NGC 3627 is likely the result of tidal interaction and \ion{H}{1} stripping by companion galaxy NGC 3628. 

NGC 3627 exhibits X-ray characteristics reflective of a galaxy that has recently undergone a starburst \citep{Dahlem96}. Intense star formation activity has been observed in the nucleus and at both ends of the galactic bar \citep{Warren10}. Low levels of star formation have also been seen in H\textsc{$\alpha$} in the western arm, while the eastern arm contains a more vigorous star-forming region in its inner section \citep{Smith94}. Higher velocity dispersions have been measured in the southern bar end relative to the northern end \citep[e.g,][]{Zhang93, Chemin03, Dumke11}. The southern bar end also has been found to exhibit double line profiles in CO \citep{Beuther17} as well as harbor an unexplained magnetic field orientation, which does not follow the underlying optical spiral arm structure \citep{Soida01}. However, in a recent CO line analysis, \citet{Watanabe11} found that the star formation rates of the southern and northern bar ends are elevated relative to all other regions of NGC 3627 but are not substantially different than one another. \citet{Beuther17} conclude that the active star formation in the bar-arm interaction regions of NGC 3627 is the result of crossing gas orbits and colliding gas clouds, piling up dense gas that can then collapse and undergo intense star formation.

\indent We present moderate (${\sim}5^{\prime \prime}$) and high angular resolution (${\sim}1^{\prime \prime}$) observations of $^{12}\rm{CO\,} (J = 2 - 1)$ and $^{13}\rm{CO\,} (J = 2 - 1)$ line emission toward NGC~3627 taken with the Submillimeter Array (SMA). These observations represent the most complete, in terms of overall spatial coverage and resolution, of $^{12}$CO ($J = 2 - 1$) emission toward NGC~3627. In Section \ref{sec:observations}, we present the SMA observations and describe the imaging process. We discuss CO morphology and kinematics and present detailed emission maps in Section \ref{sec:CO_Emission_section}. In Section \ref{sec:13CO_section}, we discuss the detection of the $^{13}\rm{CO\,} (J = 2 - 1)$ isotopologue and then we use RADEX modeling to derive physical properties of the molecular gas in Section \ref{sec:modeling_section}. We derive a galactic rotation curve and dynamical mass estimates in Section \ref{sec:rotation_curve_section} and summarize our results in Section \ref{sec:conclusions}.

\begin{deluxetable}{lcccc}
\tablecaption{Basic Astronomical Properties of NGC\,3627\label{tab:basic_info}}
\tablehead{\colhead{Property} & \colhead{Value} & \colhead{Ref.}}\vspace{-0.3cm}
\startdata
R.A. (J2000) & $11^h20^m15.02^s$ & 1 \\[-0.1cm]
Dec. (J2000) & $+12^{\circ}59^{\prime}29^{\prime \prime}.50$ & 1\\[-0.1cm]
Classification & SAB(s)b & 2 \\[-0.1cm]
Arm Class & Grand Design Spiral (7) & 3\\[-0.1cm]
Nucleus & LINER/Sy2 & 4\\[-0.1cm]
Distance (Mpc) & $10.57 \pm 0.73$ & 5 \\[-0.1cm]
Linear Scale (pc arcsec$^{-1}$) & 51 & 5\\[-0.1cm]
Inclination ($^{\circ}$) 	& $61.3$ & 1 \\[-0.1cm]
Position Angle ($^{\circ}$) & $178.0 \pm 1$ & 1 \\[-0.1cm]
V$_{\rm{hel}}$ ($\rm{km}\,\rm{s}^{-1}$) & $744$ & 1 \\[-0.1cm]
Stellar Mass ($M_{\odot}$) & $10.23 \times 10^{10}$ & 6\\[-0.1cm]
Orbital Mass ($M_{\odot}$) & $(1.45 \pm 0.39) \times 10^{12}$ & 6 \\[-0.1cm]
H$_2$ Mass ($M_{\odot}$) & $9.6 \times 10^{9}$ & 7 \\
\enddata
\tablecomments{(1) \citet{Casasola11}; (2) \citet{Vaucouleurs91}; (3) \citet{Elmegreen87}; (4) \citet{Ho97}, (5) \citet{Lee13}; (6) \citet{Karachentsev14}; (7) This work.}
\end{deluxetable}

\section{Observations}
\label{sec:observations}
\subsection{Submilllimeter Array Observations}

NGC 3627 was observed with the SMA\footnote{The Submillimeter Array is a joint project between the Smithsonian Astrophysical Observatory and the Academia Sinica Institute of Astronomy and Astrophysics and is funded by the Smithsonian Institution and the Academia Sinica.} \citep{Ho04} between 2016 March 27 and 2017 May 31 in the sub-compact, compact, and extended configurations. All observations contained 7 or 8 antennas and covered a range of baseline lengths from $9.5$--$226.0$~m. Observations consisted of a 32-point mosaic centered at R.A.(J2000) $=$11$^{\rm{h}}$20$^{\rm{m}}$15$^{\rm{s}}$ and Dec.(J2000)$=+$12$^{\circ}$59$^{\prime}$30$^{\prime \prime}$ with half-beam spacing. In order to improve \textit{uv}-coverage, these observations were repeated in different antenna configurations. The LO frequency was 224.92 GHz, which placed the $^{12}$CO ($J=2-1$) line in the upper sideband, and the $^{13}$CO ($J=2-1$) and C$^{18}$O ($J=2-1$) lines in the lower sideband. For all observations, J1058$+$015 and 3c274 were used as the primary phase and amplitude gain calibrators with absolute flux calibration performed by comparison to Callisto. Passband calibration used 3c273 for all observations, except 2017 Feb 6, which used 3c84. 

The system temperature ranged from ${\sim}140\,$--$\,230 \rm{\,K}$ depending on source elevation. At 1.3 mm, the SMA primary beam is ${\sim}55^{\prime \prime}$ (FWHP), and the largest recoverable scales for the array in the sub-compact, compact, and extended configurations are ${\sim}28^{\prime \prime}$, ${\sim}16^{\prime \prime}$, and ${\sim}6^{\prime \prime}$, respectively. The total observed bandwidth ranged from 8 to 32 GHz, as these data were taken during the commissioning of a new correlator. Depending on the SMA correlator that was used, spectral resolution was either 0.8125 MHz or 140.0 kHz per channel. A complete summary of observations is given in Table \ref{tab:SMA_obs}.

\begin{deluxetable*}{lccccccc}[!htp]
\caption{SMA Observations\label{tab:SMA_obs}}
\tablehead{[-.3cm]
\colhead{UT Date} & \colhead{Config.} & \colhead{Number} & \colhead{Baseline Range} & \colhead{$\tau$} & \multicolumn{3}{c}{Calibrators} \\
\cline{6-8} \\ [-.5cm]
& & \colhead{Antennas} & \colhead{(m)} & \colhead{(225 GHz)} & \colhead{Flux} & \colhead{Passband} & \colhead{Gain} \\[-.6cm]}
\startdata
2016 Mar 27 & Compact        & 7 & 20.3--81.4 & 0.05--0.07  & Callisto & 3c273 & J1058$+$015, 3c274 \\[-.15cm]
2016 Mar 28 & Compact        & 8 & 16.4--81.4 & 0.06           & Callisto & 3c273 & J1058$+$015, 3c274 \\[-.15cm]
2016 Mar 29 & Compact        & 8 & 16.4--81.4 & 0.04--0.07  & Callisto & 3c273 & J1058$+$015, 3c274 \\[-.15cm]
2016 Apr 22 & Extended       & 8 &44.2--226.0 & 0.02--0.06 & Callisto & 3c273 & J1058$+$015, 3c274 \\[-.15cm]
2017 Feb 6   & Sub-compact  & 7& 9.5--68.4 & 0.10--0.15     & Callisto & 3c84  & J1058$+$015, 3c274  \\[-.15cm]
2017 May 31 & Sub-compact & 7 & 9.5--45.2 & 0.13--0.15    & Callisto & 3c273 & J1058$+$015, 3c274  \\
\enddata
\end{deluxetable*}

\begin{figure}[ht]
\centerline{%
\includegraphics[scale=0.4]{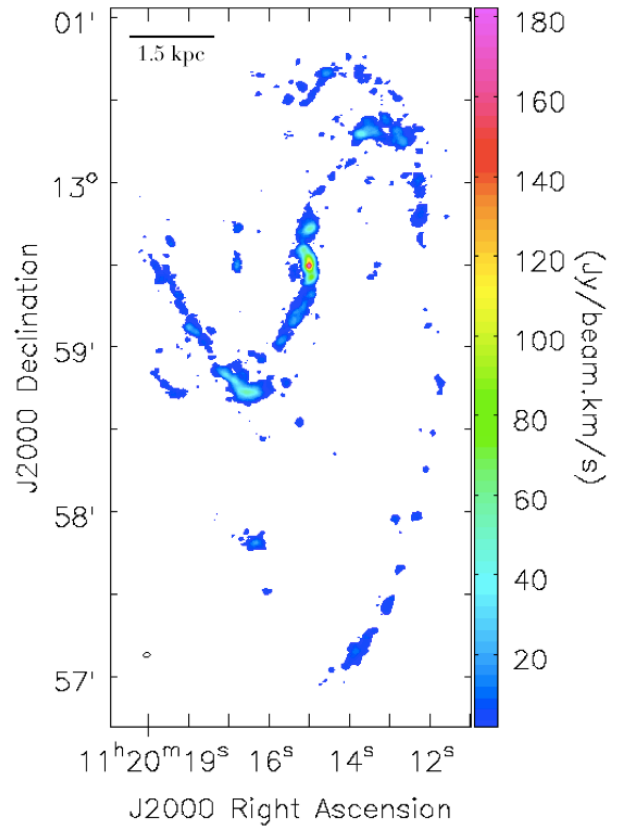}%
}%
\caption{SMA moment-0 map of the $^{12}\rm{CO\,}(J=2-1)$ emission in NGC~3627 imaged with natural weighting and no \textit{uv}-taper. The synthesized beam is shown in the lower left corner with FWHM size of $2.25^{\prime \prime} \times 1.75^{\prime \prime}$ and position angle of 95.82$^{\circ}$.} 
\label{fig:moments_1and2}
\end{figure}

\begin{figure*}[ht]
\centerline{%
\includegraphics[scale=0.7]{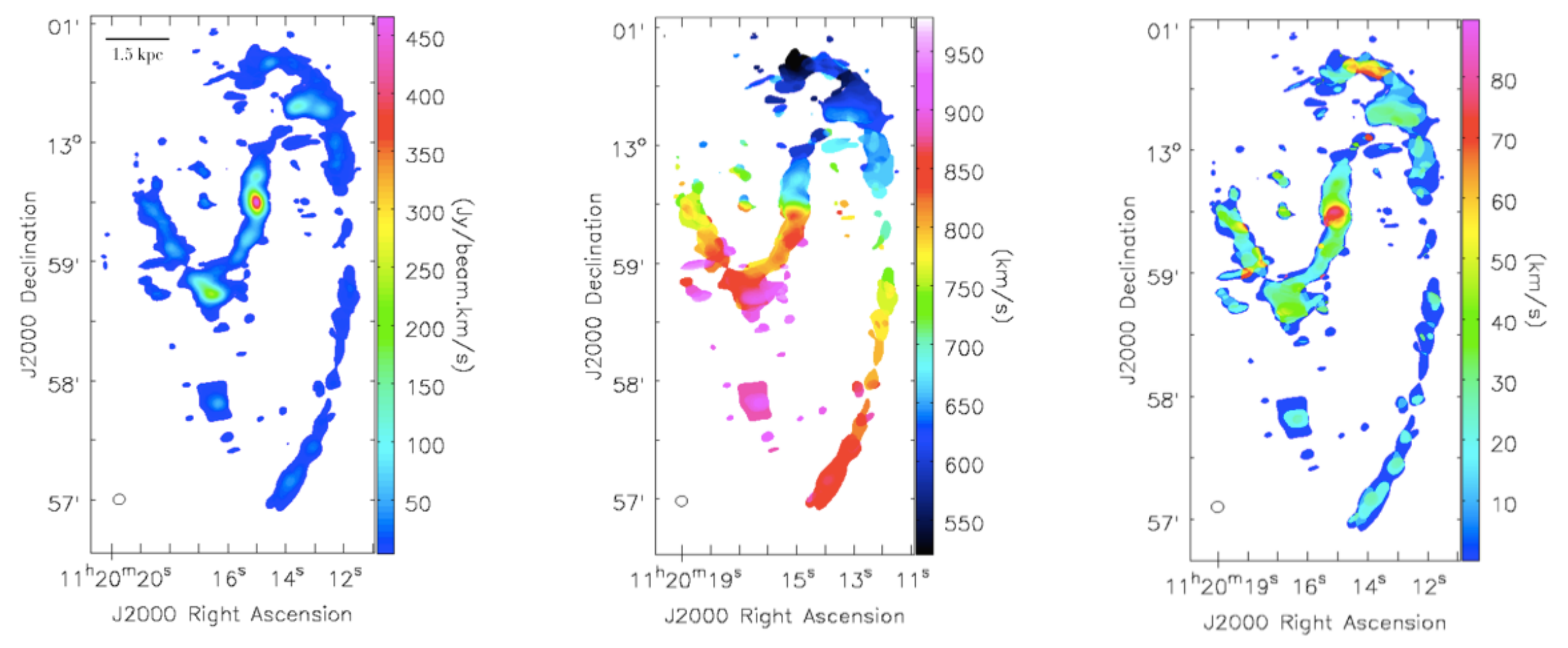}%
}%
\caption{\textit{Left}: SMA moment-0 map of the $^{12}\rm{CO\,}(J=2-1)$ emission in NGC~3627 imaged with natural weighting and an outer \textit{uv}-taper of 6 arcsec. The synthesized beam is shown in the lower left corner with FWHM size of $6.07^{\prime \prime} \times 5.22^{\prime \prime}$ and position angle of 93.29$^{\circ}$. \textit{Middle}: SMA moment-1 map imaged as in left panel. A substantial velocity gradient is seen in the central bar as well as between the north and south ends of the bar. \textit{Right}: SMA moment-2 map imaged as in left panel. The highest velocity dispersion is found in the central bar with the next largest dispersion being seen in the southern end of the galactic bar.} 
\label{fig:moment_0s}
\end{figure*}

The SMA data were calibrated using the MIR software package\footnote{http://www.cfa.harvard.edu/${\sim}$cqi/mircook.html}. The calibrated visibilities were then exported into CASA \citep{McMullin07} and the task \texttt{tclean} was used to Fourier transform the complex visibilities to create an image of the $^{12}$CO$(J=2-1)$ emission. The dirty image revealed prominent regions of line emission, which occur primarily in the nuclear bar and two extended spiral arms, and we used this information to supply an appropriate mask of width ${\sim}20$--$30^{\prime \prime}$ tracing these regions. We adopt multi-scale cleaning, allowing model Gaussian components of width 0$^{\prime \prime}$ (delta function), 6$^{\prime \prime}$,  15$^{\prime \prime}$, and 30$^{\prime \prime}$. Figure \ref{fig:moments_1and2} presents an SMA map of the $^{12}$CO$(J=2-1)$ emission in NGC~3627 obtained with natural weighting using a Briggs robust parameter of 2. This weighting scheme gives a synthesized beam of $2.25^{\prime \prime} \times 1.75^{\prime \prime}$ with PA$=95.82^{\circ}$, corresponding to a spatial resolution of about $115 \times 89\,\rm{pc}$ at the distance of NGC 3627. The leftmost panel of Figure \ref{fig:moment_0s} shows the same SMA data but weighted with an additional Gaussian \textit{uv}-taper with an on-sky FWHM of 6$^{\prime \prime}$. The taper was used to suppress the weight of the longer baselines and highlight the more diffuse, extended emission in the spiral arms of NGC 3627. The angular resolution of the map is 6.07 $^{\prime \prime} \times$ 5.22$^{\prime \prime}$ with PA$=$93.29$^{\circ}$, which corresponds to an approximate physical size of $310 \times 266\,\rm{pc}$ and is approximately 8$\times$ larger than the beam size obtained with natural weighting alone. All maps were binned into 20 km s$^{-1}$ velocity channels to enhance signal-to-noise. The typical $1\sigma$ rms noise levels per individual mosaic field on the natural-weighted and tapered maps are 14 and 20 $\rm{mJy}\,\rm{beam}^{-1}\,\rm{channel}^{-1}$, respectively. No conspicuous 230~GHz continuum emission was found in NGC 3627 down to a $3\sigma$ level in either the natural-weighted or tapered maps.

\section{Results and Discussion}
\subsection{CO Emission and Morphology}
\label{sec:CO_Emission_section}
\subsubsection{Large-scale CO Distribution and Kinematics}
As seen in the integrated intensity maps in Figures \ref{fig:moments_1and2} and \ref{fig:moment_0s}, strong $^{12}$CO$(J=2-1)$ emission is detected in the central bar with two prominent emission regions located at the northern and southern regions of the bar, $51^{\prime \prime}$ and $52^{\prime \prime}$ from the nuclear region, respectively. Emission extending $16^{\prime \prime}$ past the northern bar region can clearly be seen. Faint emission is also present in the more extended western spiral arm (${\sim}220^{\prime \prime}$ in length) and in the smaller eastern arm (${\sim}65^{\prime \prime}$ in length). A relatively isolated clump of emission (hereafter ``the clump") located ${\sim}55^{\prime \prime}$ to the south of the southern end of the bar is also clearly detected. 

The 1st and 2nd moment maps, corresponding to intensity-weighted peak velocities and velocity dispersions, respectively, are presented in Figure \ref{fig:moment_0s} and reveal substantial kinematic structure. Velocity gradients of similar magnitudes (${\sim}250\,\rm{km}\,\rm{s}^{-1}$) are present throughout the central bar as well as between the northern and southern regions of the bar. The spiral arms, particularly the extended western arm, display the largest gradients with velocities ranging from 500--900 km s$^{-1}$. The broadest lines, as indicated by the moment-2 map, are found toward the integrated intensity peaks, namely the nuclear bar (${\sim}60$--80 km s$^{-1}$), the southern end of the bar (${\sim}35$--50 km s$^{-1}$), and the northern end of the bar (${\sim}20$--35 km s$^{-1}$).  Velocity dispersions are typically narrow in the more extended spiral arms with typical dispersions of $\lesssim15$ km s$^{-1}$. The region of emission to the north of the northern bar end has particularly broad lines of up to ${\sim}70$ km s$^{-1}$.

\begin{figure}[ht]
\centering
\includegraphics[scale=0.4]{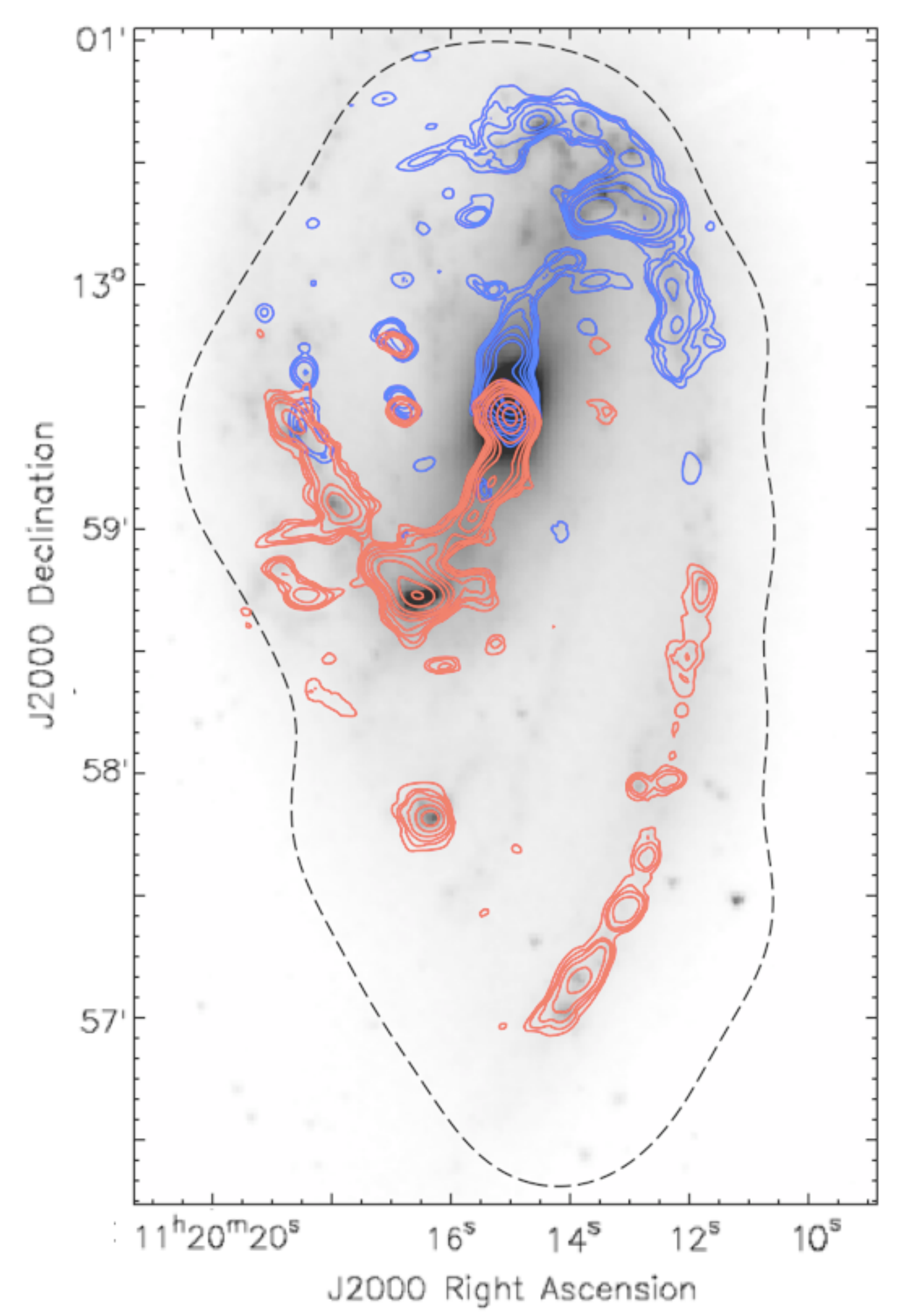}
\caption{SMA $^{12}\rm{CO\,}(J=2-1)$ integrated-intensity contour overlay of Spitzer/IRAC 3.6 $\mu$m image \citep{Kennicutt03}. The red and blue contours are with respect to $v_{\rm{hel}} = 744$ km s$^{-1}$ and represent $\left(2, 3, 4, 6, 10, 15, 25, 50, 75, 85 \right) \times \sigma = 3.0 \rm{\,Jy\,}\rm{km}\,\rm{s}^{-1}$ of the image made with natural weighting and a 6 arcsec \textit{uv}-taper at an angular resolution of $6.07^{\prime \prime} \times 5.22^{\prime \prime}$. There is reasonable agreement between the CO emission and infrared image. The black dashed contours indicate the spatial coverage up to 30\% of the primary beam response of the mosaiced image.}
\label{fig:cuts_infrared_images}
\end{figure}

Figure \ref{fig:cuts_infrared_images} presents our SMA $^{12}$CO($J=2-1$) emission overlaid on a \textit{Spitzer} 3.6 $\mu$m image, which traces the stellar component and is available due to the \textit{Spitzer} Infrared Nearby Galaxies Survey project \citep[SINGS;][]{Kennicutt03}.  Over the whole galaxy, the molecular gas roughly traces the stellar component. The contours, which indicate galactic rotation, are with respect to a systemic velocity of 744 km s$^{-1}$ \citep{Casasola11}. The spatial coverage up to 30\% of the primary beam response of the SMA mosaic is shown as a black dashed line.

\subsubsection{Nuclear Region}
A strong nuclear H\textsc{$\alpha$} emission line initially identified the nucleus of NGC 3627 as undergoing some form of weak starburst \citep{Filippenko85}. More recent high-resolution maps of the $^{12}$CO($J=1-0$) and $^{12}$CO($J=2-1$) lines in the nuclear region of NGC 3627 by \citet{Casasola11} reveal an ${\sim}18^{\prime \prime}$ bar-like inner structure with a PA=$14^{\circ}$ with two emission peaks at either ends of an elongated emission region. They also found that the stellar bar, traced by 1.6 $\mu$m and 3.6 $\mu$m \textit{Spitzer} images, has a PA=$-21^{\circ}$, indicating that the gas is leading the stellar bar.

\begin{figure*}[htp!]
\centering
\includegraphics[width=\textwidth]{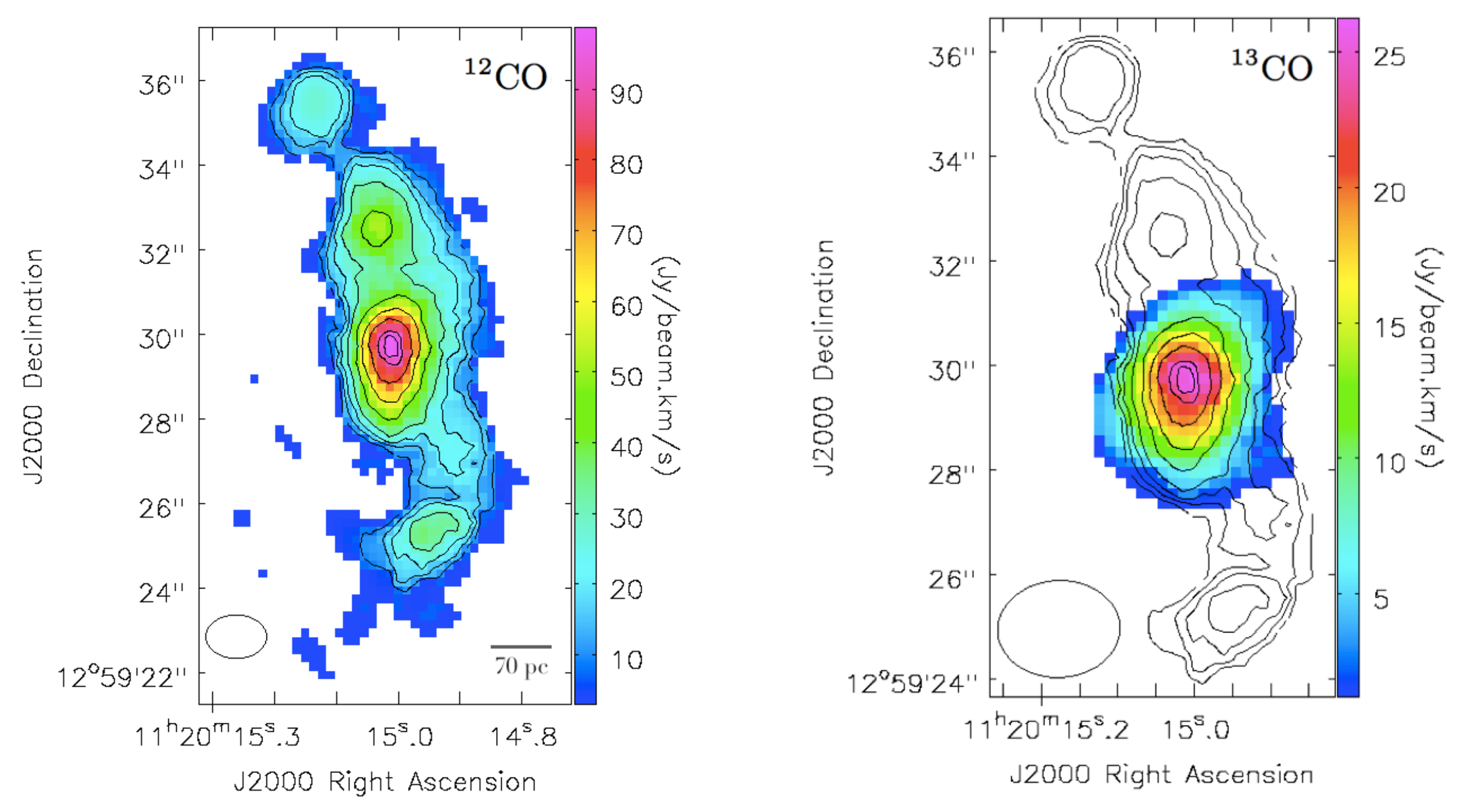}
\caption{\textit{Left}: SMA moment-0 map of the nuclear region of NGC 3627 in $^{12}$CO($J=2-1$) emission made with uniform weighting. The contours represent $\left(3,5,7,10,15,20,25,30,32\right) \times \sigma = 3.0$ Jy km s$^{-1}$. The synthesized beam is shown in the lower left corner with FWHM size of $1.41^{\prime \prime} \times 1.01^{\prime \prime}$ with position angle of $88.77^{\circ}$. \textit{Right}: The $^{13}$CO($J=2-1$) emission is shown in color scales and $^{12}$CO($J=2-1$) in contours. The synthesized beam is shown in the lower left corner with FWHM size of $2.33^{\prime \prime} \times 1.85^{\prime \prime}$ with position angle of $92.60^{\circ}$.}
\label{fig:13CO_center_img}
\end{figure*}

We imaged the central region of NGC 3627 using uniform weighting ($r=-2$) to highlight small-scale features and a rectangular mask of width ${\sim}5^{\prime \prime}$ centered on the SMA phase center of NGC 3627, which resulted in a beam size of $1.41^{\prime \prime} \times 1.01^{\prime \prime}$ with PA $=88.77^{\circ}$ and a typical rms noise level of $35$\,mJy beam$^{-1}$ per channel. The central $^{12}$CO($J=2-1$) emission is shown in the left panel of Figure \ref{fig:13CO_center_img}. While our beam size for the uniform weighting image is about 2.5 times larger than the high-resolution study made by \citet{Casasola11}, who had a $^{12}$CO($J=2-1$) synthesized beam of $0.9^{\prime \prime} \times 0.6^{\prime \prime}$ (see their Figure 4), we successfully recover small-scale structures, namely two emission maxima offset from the central peak, as reported in \citet{Casasola11}. 

\subsection{Detection of $^{13}$CO($J=2-1$) and Line Ratios}
\label{sec:13CO_section}
\begin{figure*}[ht]
\centering
\includegraphics[width=\textwidth]{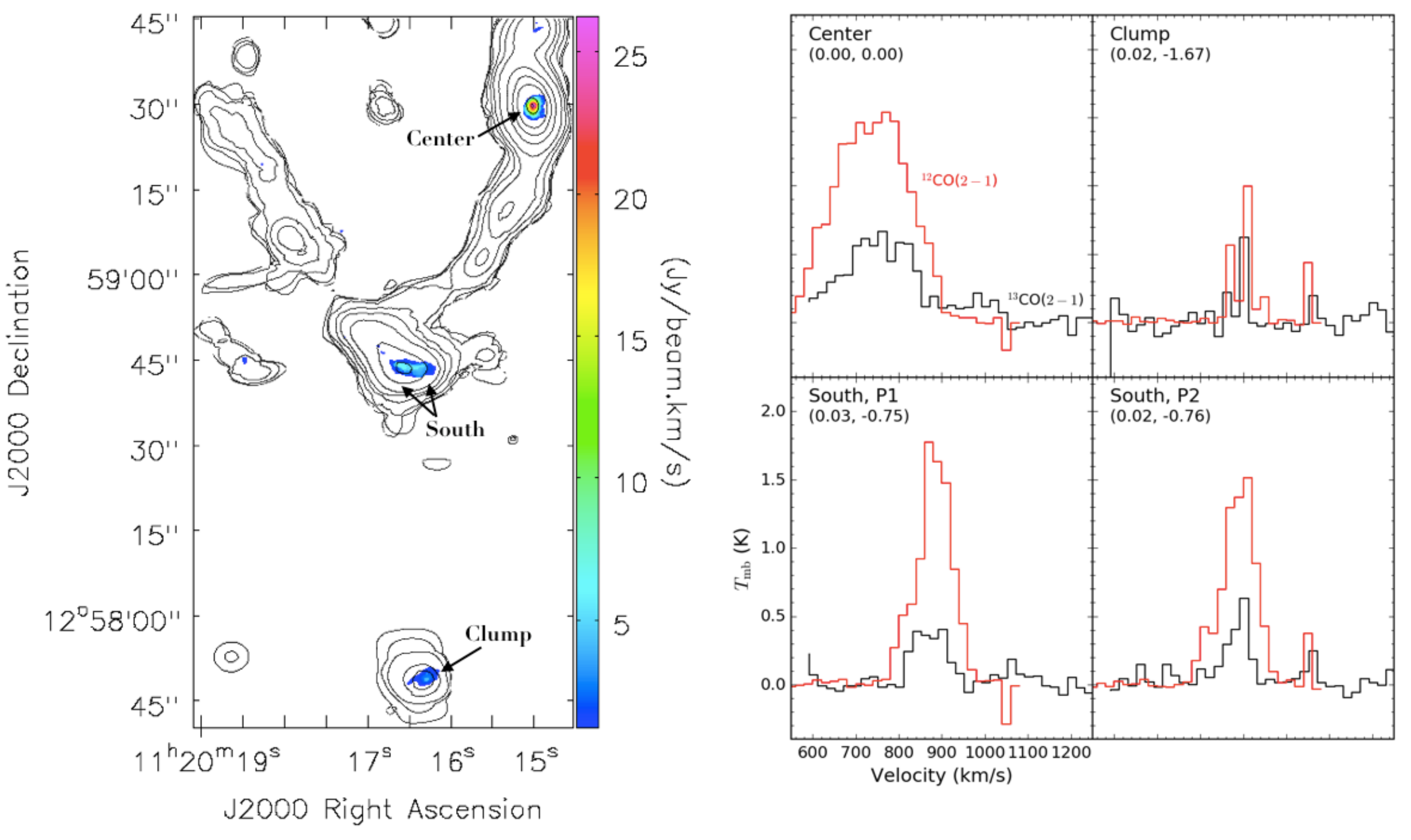}
\caption{\textit{Left}: SMA $^{13}\rm{CO\,}(J=2-1)$ integrated-intensity moment-0 map overlaid on $^{12}\rm{CO\,}(J=2-1)$ contours. The black contours are from the image made with a 6 arcsec \textit{uv}-taper and represent $\left( 2, 3, 5, 10, 15, 25, 50, 75, 100, 145, 155 \right) \times \sigma = 3.0 \rm{\,Jy\,}\rm{km}\,\rm{s}^{-1}$. The inner beam indicates the angular resolution of the $^{13}$CO($J=2-1$) emission, which is $2.33^{\prime \prime}\times1.85^{\prime \prime}$ with PA$=92.60^{\circ}$ and $1\sigma$ rms of $14$~mJy beam$^{-1}$ channel $^{-1}$, and the outer beam shows the resolution of the \textit{uv}-tapered $^{12}$CO($J=2-1$) emission. \textit{Right}: Spectra of $^{12}$CO($J=2-1$)  in red and $^{13}$CO($J=2-1$) in black from the regions where $^{13}$CO($J=2-1$) spectra could be extracted. Offsets in arcminutes from the center position are shown in the upper left corner.} 
\label{fig:13CO_with_whole_cont}
\end{figure*}

Emission from $^{13}$CO($J=2-1$) was also detected in NGC 3627. The $^{13}$CO($J=2-1$) line is generally optically thin and traces the molecular gas mass more accurately than the $^{12}$CO($J=2-1$) emission, which often arises from the surface of molecular clouds. The $^{13}$CO($J=2-1$) emission was imaged in the same way as described in Section \ref{sec:observations} with natural weighting, which resulted in a beam size of $2.33^{\prime \prime}\times1.85^{\prime \prime}$ with PA$=92.60^{\circ}$ and a typical rms noise level of $14$\,mJy beam$^{-1}$ channel $^{-1}$. Figure \ref{fig:13CO_with_whole_cont} shows the spatial distribution of detected $^{13}$CO($J=2-1$) relative to the more extended $^{12}$CO($J=2-1$) emission. The most prominent $^{13}$CO($J=2-1$) emission is found in the nuclear region of NGC 3627, but additional emission is observed in the southern bar end and the clump. No $^{13}$CO($J=2-1$) emission was detected around the northern bar end or in the more diffuse spiral arms. None of the $^{13}$CO($J=2-1$) emission was spatially resolved and no emission from the C$^{18}$O($J=2-1$) line in NGC~3627 was detected down to a $3\sigma$ rms noise level of $42$~mJy beam$^{-1}$ per channel.

The right panel of Figure \ref{fig:13CO_center_img} shows a zoomed-in image of the $^{13}$CO($J=2-1$) emission in the central region of NGC~3627. While we do not resolve any spatial structure, we note that the nuclear $^{13}$CO($J=2-1$) emission is coincident with the maximum peak in the $^{12}$CO($J=2-1$) line. 

Despite the limited detections of the $^{13}$CO($J=2-1$) line, we were able to investigate the integrated-intensity line ratios at four different positions in the galaxy, as shown in the left panel of Figure \ref{fig:13CO_with_whole_cont}. We extracted spectra averaged in an area equal to the beam size of the $^{13}$CO($J=2-1$) emission, i.e. $2.33^{\prime \prime} \times 1.85^{\prime \prime}$, in the center, south, and clump for both $^{13}$CO($J=2-1$) and $^{12}$CO($J=2-1$) transitions. The spectra were fit with single Gaussian profiles and the line ratios are reported in Table \ref{tab:13CO_line_ratios_tab}. While \citet{Tan11} find an elevated $R_{12/13}$ line ratio of $10.4\pm3.6$ for the $J=1-0$ transition in the central region of NGC 3627, we find a ratio of $2.51 \pm 0.23$, about one-fifth of this value, in the $J=2-1$ transition. The $R_{12/13}$ values in the southern bar end and the clump are also higher (${\sim}30\%$) than those for the $J=1-0$ transition. Similarly, \citet{Cormier18} report $J=1-0$ ratios in excess of 10 for all galactocentric radii $\lesssim\,9$~kpc, reporting a central $R_{12/13}$ of $15.2\pm0.5$.

Since we are comparing our isotopic ratios for the $J=2-1$ transition with that of lower resolution observations of the $J=1-0$ line, the observed differences may be due to a beam-filling factor effect. It is possible that $^{12}$CO is much more diffuse and easily detectable everywhere (in the large beams), but the $^{13}$CO, which also probes intermediate density gas, may be more compact, which would tend to make the low resolution ratios appear larger. However, when we resolve structure, as in the SMA maps, and select regions where the emission fills the beam, the ratios would appear smaller.

\begin{deluxetable*}{lccccccccc}[!htp]
\caption{$^{12}$CO($J=2-1$) / $^{13}$CO($J=2-1$) Integrated-Intensity Ratios \label{tab:13CO_line_ratios_tab}}
\tablehead{[-.3cm]
\colhead{$\Delta \alpha$} & \colhead{$\Delta \delta$} &\colhead{Location} &\colhead{$I_{^{12}\rm{CO}}$}&\colhead{$V_{\rm{c}}$} &\colhead{$\Delta V$}& \colhead{$I_{^{13}\rm{CO}}$} & \colhead{$V_{\rm{c}}$} & \colhead{$\Delta V$} & $R_{12/13}$\\[-.2cm]
\colhead{($^{\prime}$)} & \colhead{($^{\prime}$)} &\colhead{} &\colhead{(K km s$^{-1}$)}&\colhead{(km s$^{-1}$)} &\colhead{(km s$^{-1}$)} & \colhead{(K km s$^{-1}$)} & \colhead{(km s$^{-1}$)} & \colhead{(km s$^{-1}$)} & \colhead{}\\[-.6cm]}
\startdata
0.00 & 0.00      & Center & $347.77 \pm 12.61$ & 748.24 $\pm$ 2.42 & $207.64 \pm 5.69$ & $138.61 \pm 11.66$ & 762.95 $\pm$ 5.77 & 219.04 $\pm$ 14.20 & 2.51 $\pm$ 0.23  \\[-.15cm]
0.02 & $-$1.67 & Clump & 40.53 $\pm$ 11.48 & 912.58 $\pm$ 5.44 & 59.84 $\pm$ 12.81 & $13.97 \pm 7.21$ & 911.29 $\pm$ 8.40 & 21.19 $\pm$ 9.37 &  2.90 $\pm$ 1.71 \\[-.15cm]
0.03 & $-$0.75 & South, P1 & 169.73 $\pm$ 9.31 & 894.96 $\pm$ 1.64 & 93.33 $\pm$ 3.87 & 40.62 $\pm$ 6.88 & 881.71 $\pm$4.90 & 90.09 $\pm$ 11.53 & 4.18 $\pm$ 0.74 \\[-.15cm]
0.02 & $-$0.76 & South, P2 & 142.54 $\pm$ 9.61 & 902.29 $\pm$ 1.95 & 90.10 $\pm$ 4.59 & 33.41 $\pm$ 6.07 & 902.14 $\pm$ 3.28 & 56.15 $\pm$ 7.72 & 4.27 $\pm$ 0.83 \\
\enddata
\end{deluxetable*}

Due to the low signal-to-noise spectral detections in the clump, the $R_{12/13}$ value in the clump has a high uncertainty (${\sim}$40\%) and is somewhat difficult to interpret, but the $R_{12/13}$ values in the nucleus and southern bar end are well-constrained with uncertainties $\lesssim$20\% and thus can be reliably compared. The denser, hotter region in the nucleus has a ratio that is ${\sim}50\%$ lower than that of the more diffuse and cooler southern bar end. The simplest physical explanation for these observed changes is that the lines may probe slightly different material, i.e. the $J=2-1$ transition has a higher critical density \citep[e.g.,][]{Shirley15} and is probing denser and warmer gas. Since both lines are likely to be similarly excited \citep{Davis14}, and in the absence of chemical effects like fractionation or nucleosynthesis, emission from the $J=2-1$ transition will occupy a smaller fraction of the beam.

Another likely driver of abundance ratios are changes in optical depth due to variations in gas physical conditions, such as gas density, temperature, and opacity \citep[e.g.,][]{Pineda08, Wong08}. By modeling the centers of active nuclei, \citet{Israel09a,Israel09b} found that different mixtures of two ISM components, one hot and tenuous with low optical depth and another that was cooler and denser, could explain the observed ranges of $R_{12/13}$ in nearby galaxies. This effect may be especially important in NGC~3627, as the presence of increased turbulence (for instance, induced by non-circular motions due to a stellar bar) lowers the optical depth of $^{12}$CO and increases that of $^{13}$CO \citep{Cormier18}. Additionally, the largest line widths, as indicated by velocity dispersion, from both $J=2-1$ and $J=1-0$ lines of $^{12}$CO in NGC~3627 are seen in the nucleus.

Other effects besides optical depths could also be influencing these ratios \citep[and see references therein]{Cormier18}. For instance, since the southern bar end has a $\Sigma_{\rm{SFR}}$ that is twice that of the nuclear region \citep{Watanabe11}, significant $^{13}$CO may have been photodissociated by strong interstellar radiation fields of newly-forming OB stars in the south. Photodissociation is known to be selective as $^{13}$CO is preferentially destroyed as the UV radiation field increases in strength, while $^{12}$CO is mostly self-shielded \citep{Dishoeck88}. However, recent simulations of molecular clouds show that selective photodissociation only has a minimal effect on the $^{12}$CO/$^{13}$CO abundance ratio \citep{Szucs14}, while \citet{Donaire17} also argue that shielding due to dust and H$_2$ is more dominant than CO self-shielding but is still insufficient to drive the observed ratios.

The differences in $R_{12/13}$ values could also be a time-scale effect due to selective nucleosynthesis of $^{12}$C and changes in $^{12}$C/$^{13}$C abundance \citep[e.g.,][]{Wilson99}. Supernovae produce conspicuous $^{12}$C but insignificant amounts of $^{13}$C, which is instead primarily produced by low-mass stars and injected into the ISM from the winds of asymptotic branch stars \citep{Sage91}. Starbursts can also lead to enhanced $^{12}$C abundances relative to $^{13}$C abundances \citep[e.g.,][]{Meier14, Sliwa17}. In fact, an increase in $^{12}$C/$^{13}$C abundance ratio with galactocentric radius is seen in the Milky Way \citep{Milam05} and has successfully been reproduced by time-dependent models \citep{Romano17}.

\subsection{Radiative Transfer Modeling}
\label{sec:modeling_section}
\subsubsection{Line Ratios Compared to BIMA Data}
\label{sec:line_ratios_compared_to_BIMA}
The relative integrated line intensities of the $^{12}$CO($J=2-1$) emission, $I_{21}$, and the $^{12}$CO($J=1-0$) emission, $I_{10}$ provide valuable information about the local excitation conditions. We acquired the $^{12}$CO($J=1-0$) data from the BIMA SONG\footnote{Berkley-Illinois-Maryland Association Survey of Nearby Galaxies} survey \citep{Regan01, Helfer03} from NED\footnote{The NASA/IPAC Extragalactic Database (NED) is operated by the Jet Propulsion Laboratory, California Institute of Technology, under contract with the National Aeronautics and Space Administration.}. The large-scale $^{12}$CO($J=1-0$) emission image was taken with the 10-element BIMA millimeter interferometer \citep{Welch96} and covers a field of $350^{\prime \prime}\times410^{\prime \prime}$.

\begin{figure}[htp!]
\centering
\includegraphics[scale=0.7]{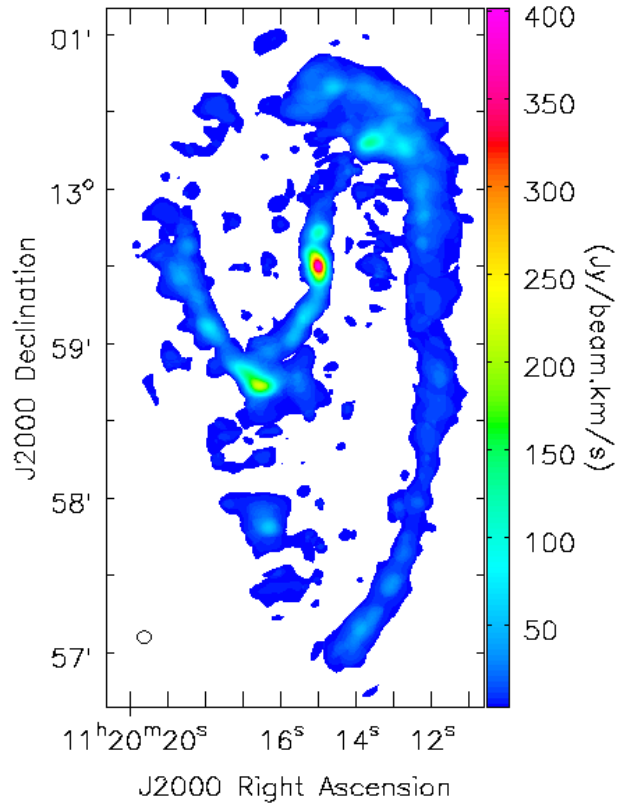}
\caption{SMA$+$IRAM~30~m combined map of the $^{12}$CO($J=2-1$) emission in NGC~3627. The synthesized beam is shown in the lower left corner with FWHM size of $5.39^{\prime \prime} \times 4.87^{\prime \prime}$ and position angle of 85.57$^{\circ}$ }
\label{fig:feather_map}
\end{figure}

The BIMA $^{12}$CO($J=1-0$) data were combined with single-dish data from the NRAO 12~m telescope on Kitt Peak, AZ \citep{Regan01}, resulting in a beam size of $7.3^{\prime \prime} \times 5.8^{\prime \prime}$ and an rms noise level of 41~mJy~beam$^{-1}$ per 10~km~s$^{-1}$ channel \citep{Helfer03}. Typical large-scale flux recovery attributable to the single-dish observations for NGC~3627 was ${\sim}10$--$30\%$ \citep{Helfer03}. In order to ensure reliable comparisons with our SMA observations, we incorporated single-dish data from the HERACLES survey taken with the IRAM 30~m telescope \citep{Leroy09} to fill in the missing short \textit{uv} spacings and remedy any ``missing flux" issues to which interferometric images are often susceptible \citep[e.g,][]{Sakamoto99}. The CASA task \texttt{feather} was used to vary the relative weights of the 30~m and SMA observations for the optimal trade-off between angular resolution and restoration of missing flux. Most recovered flux was concentrated in the extended regions of NGC~3627 including the inter-arm regions and spiral arms with flux recovery ranging from $40\%$--$70\%$ in the most diffuse regions. The SMA$+$IRAM~30~m combined map, which is shown in Figure \ref{fig:feather_map}, has a resolution of $5.39^{\prime \prime} \times 4.87^{\prime \prime}$ and an rms noise level of 30~mJy~beam$^{-1}$ per 20~km~s$^{-1}$ channel.

\begin{figure}[htp!]
\centering
\includegraphics[scale=0.5]{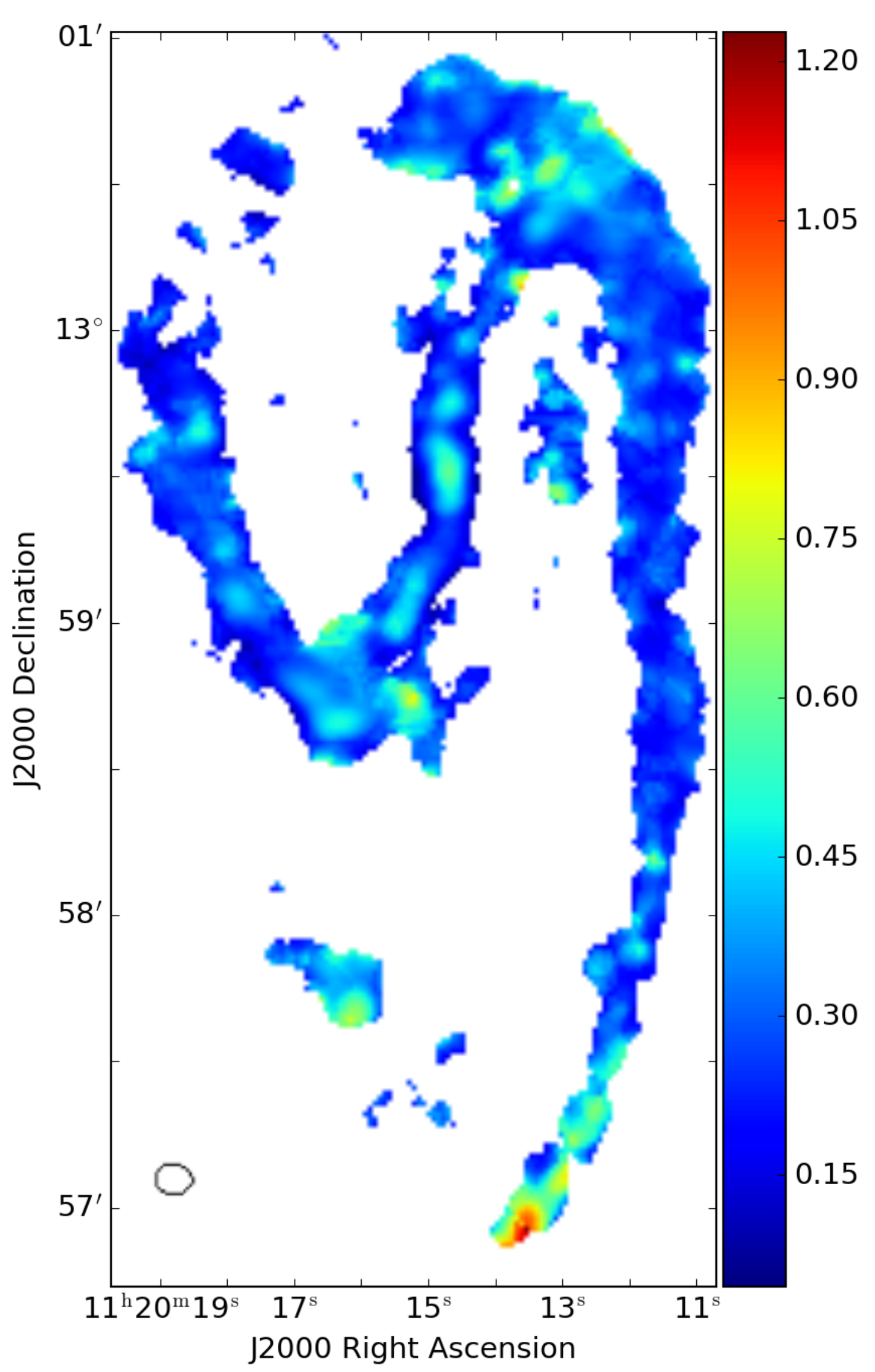}
\caption{Line ratio map of $^{12}$CO($J=2-1$) / $^{12}$CO($J=1-0$) with a $3\sigma$ threshold. The synthesized beam is shown in the lower left corner with FWHM size of $7.27^{\prime \prime} \times 5.77^{\prime \prime}$ and position angle of $80.19^{\circ}$. Combined SMA$+$IRAM~30~m and BIMA$+$NRAO~12~m data were used for the $^{12}$CO($J=2-1$) and $J=1-0$ emission, respectively. Moderate ratio gas (${\sim}$0.4--0.6) is seen in the nuclear region and bar ends, while the spiral arms exhibit lower ratio gas (${\sim}$0.2). Moderate-to-high gas ratios are seen in the southern portions of the extended western arm (${\gtrsim 1.0}$) and the clump (${\sim}$0.6--0.7), possibly due to recent tidal interaction with NGC~3628. Hot spots of moderate ratio gas are seen throughout NGC~3627 and appear especially prominent in the spiral arms and inter-arm regions.}
\label{fig:line_ratio_3sigma}
\end{figure}

To compute the line ratio $R_{21/10} = I_{21}$/$I_{10}$, we convolved our SMA data to the same resolution of the BIMA data. After both maps were each clipped at the $3\sigma$ level, we computed ratios for pixels with mutual $^{12}$CO emission. The resulting line ratio map is presented in Figure \ref{fig:line_ratio_3sigma}. The nuclear region exhibits typical line ratios of ${\sim}0.5$--$0.6$ over a relatively large physical scale, corresponding to the nuclear bar shown in Figure \ref{fig:13CO_center_img}. The southern bar end has slightly lower ratios of ${\sim}0.4$--$0.5$, while the northern end exhibits lower but comparable ratios of ${\sim}0.4$. The spiral arms contain primarily low-ratio gas of ${\sim}$0.2--0.3 with some of the more diffuse regions in the extended western arm having ratios as low as ${\sim}0.15$. The line ratios we report are consistent with previous studies, such as \citet{Casasola11} who found that the bulk of nuclear emission was between 0.4--0.7 and with $R_{21/10}=0.6$ reported by \citet{Krips08}. We also note that the ratio map is not homogeneous with hot spots of elevated gas ratios appearing throughout the galaxy. For instance, we see ${\sim}3^{\prime \prime}$ clumps exhibiting ratios from ${\sim}$0.7--0.8 in the southern bar end and of ${\sim}$0.6 in the northern bar end, as well as numerous smaller, ${\sim}2^{\prime \prime}$ enhancements with typical ratios between ${\sim}$0.4--0.6 in the spiral arms and the inter-arm regions.

In general, as the galactocentric radius increases, $R_{21/10}$ decreases with two notable exceptions: the clump, which contains some moderate ratio gas (${\sim}0.6$--$0.7$) and the southernmost region of the extended western arm, which exhibits the highest ratio gas seen in NGC~3627. In the clump, we observe a sharp south-to-north transition with the southernmost gas having a moderate ratio of ${\sim}0.75$ and the northernmost gas having extremely low ratio gas of ${\sim}0.2$. In the southern end of the western arm, we find ratios as high as 1.2 and typical ratios of ${\sim}0.6$--$0.8$.

The high-to-moderate line ratios seen in the nuclear region and bar ends are likely the result of nuclear starburst \cite[e.g.,][]{Filippenko85, Warren10, Casasola11} and bar-arm interactions \citep[e.g.,][]{Beuther17}, respectively. However, the elevated ratios seen in the southern portions of the clump and in the extended western arm are more surprising. These enhancements in line ratios correspond to hot spots in the \textit{Spitzer} 8$\mu$m map \citep{Kennicutt03} and have been identified as regions of active star formation \citep{Warren10}. Considering the interaction history of NGC~3627, these higher ratios are likely indicative of warm and dense molecular material triggered by tidal interaction with neighboring galaxy NGC~3628. Similar high-ratio $^{12}$CO gas has been observed by \citet{Muller14} in the tidally perturbed tail of the Small Magellanic Cloud. While these hotspots in the southern tip of the western arm and clump do not appear as significant enhancements in physical conditions, especially $n_{\rm{H}_2}$ (see the RADEX analysis in Section \ref{sec:TK_nH2_RADEX}), we believe that this is a line width effect, i.e. the low line widths ($<20$~km~s$^{-1}$) in the outer spiral arms tend to lower the estimations of temperature and density, but these regions of high-excitation gas are nonetheless worthy of follow-up investigation.

\subsubsection{Kinetic Temperature and Density} \label{sec:TK_nH2_RADEX}

To estimate the physical properties of the molecular gas, we use a large velocity gradient (LVG) radiative transfer model \citep[e.g.,][]{Scoville74, Goldreich74} to constrain excitation conditions throughout NGC~3627. We employ the RADEX code \citep{vanderTak07} with the \texttt{myRadex}\footnote{\url{https://github.com/fjdu/myRadex}} solver for non-local-thermodynamic-equilibrium (non-LTE) conditions to generate a two-dimensional parameter grid with regularly-spaced kinetic temperatures ($T_{\rm{K}}$) and H$_2$ number densities ($n_{\rm{H}_2}$). We assume a uniform spherical geometry and fix the velocity gradient to a typical value of 10 km s$^{-1}$ pc$^{-1}$, since we noticed that both our kinetic temperatures and $n_{\rm{H}_2}$ values were relatively insensitive to the choice of velocity gradient over a range of ${\sim}$1.0--60 km s$^{-1}$ pc$^{-1}$. Specifically, we generated CO intensities from models in logarithmic steps of 0.1 with kinetic temperatures ranging from $10^{0.5}$~K to $10^{2.5}$~K and H$_2$ number densities between $10^{2}$~cm$^{-3}$ and $10^{5}$~cm$^{-3}$. The CO abundance is assumed to be $10^{-4}$ \citep{Blake87} so we are sampling CO column densities between $10^{16.1}$ cm$^{-2}$ and $10^{21.6}$ cm$^{-2}$, which are consistent with typical galactic values \citep[e.g.,][]{Pety13, Muraoka16, Hunt17}. All parameters were calculated assuming a cosmic microwave background radiation temperature of $T=2.73$~K. We have assumed a single velocity component and a beam filling factor of 1 throughout the analysis. The collision rates between CO and ortho-/para-H$_2$ are extracted from LAMDA \citep{Yang10} and the ortho-/para-H$_2$ ratios are calculated based on kinetic temperatures. We adopt a 10\% uncertainty in the SMA and BIMA fluxes.

To incorporate additional kinematic information from our observations, we use the FWHM line widths taken from the SMA moment-2 map (i.e, FWHM${\sim}2.35\times\sigma$) to generate additional two-dimensional grids of 8 different velocity dispersions ranging from 5 km s$^{-1}$ to 75 km s$^{-1}$ in steps of 10 km s$^{-1}$ with each pixel being assigned to the nearest grid based on these intervals. Each grid velocity dispersion was treated as a representative midpoint with a range of $\pm 5$ km s$^{-1}$, e.g., a pixel with velocity dispersion of 19 km s$^{-1}$ would be assigned to a grid created with a $\sigma$ of 15 km s$^{-1}$. Pixels in the spiral arms and inter-arm regions that lacked SMA line widths and had been included from single-dish observations were assigned to the $\sigma=15$~km s$^{-1}$ grid, typical of the line widths seen in these more diffuse regions (e.g., rightmost panel of Figure \ref{fig:moment_0s}).

We then used a customized routine \citep{Zhang14, Lu17} to calculate the most likely beam-averaged gas kinetic temperatures and H$_2$ number densities to reproduce the observed $^{12}$CO ($J=2-1$) and $^{12}$CO($J=1-0$) integrated intensities as well as the observed $R_{21/10}$ line ratios. For each individual model, a $\chi^2$ value was calculated from the differences in the ratios of observed line brightness temperatures and those calculated from the LVG model. This process was then repeated for each pixel in the line ratio map shown in Figure \ref{fig:line_ratio_3sigma}. Further details of this procedure can be found in \citet{Zhang14} and an example of its application is shown in Appendix A in \citet{Lu17}. 

Figure \ref{fig:RADEX_maps} shows maps of the kinetic temperature and $n_{\rm{H}_2}$ derived from RADEX modeling. The kinetic temperatures range from ${\sim}5$--$10$~K in the diffuse spiral arms to ${\sim}25$~K in the nuclear region of NGC~3627. Elevated temperatures of ${\sim}15$--$20$~K are also found in the southern and northern bar ends. The derived $n_{\rm{H}_2}$ values exhibit a similar trend, spanning more than an order of magnitude from ${\sim}400$--$1000\,\rm{cm}^{-3}$ in the spiral arms to ${\sim}12500~\rm{cm}^{-3}$ in the nuclear region. The values for $T_{\rm{K}}$ are consistent with those derived using CO line ratios and an LVG approximation for other external galaxies, e.g., $10$--$30$~K in the barred spiral galaxy NGC~2903, $20$--$30$~K in the giant \ion{H}{2} region NGC~604 in M~33 \citep{Muraoka12, Muraoka16}. Our kinetic temperature estimates are also roughly consistent with those derived by \citet{Tan11}, who report typical values of ${\sim}20$ K in the northern and southern bar ends and a peak value of 44~K in the center of NGC~3627, which is about ${\sim}55\%$ higher than our nuclear estimate. We also find $n_{\rm{H}_2}$ values, especially those that correspond to the center and bar ends of NGC~3627, that are an order of magnitude in excess of those reported for NGC~2903 \citep[ $1000$--$3700~\rm{cm}^{-3}$;][]{Muraoka16} and NGC~604 \citep[$800$--$2500~\rm{cm}^{-3}$;][]{Muraoka12}.

The temperatures and densities derived from RADEX have uncertainties of ${\sim}50\%$, which include systematic errors from assumed CO abundance and column densities and random errors from observed fluxes and line widths. Determinations of $T_{\rm{K}}$ were significantly more insensitive to uncertainties in fluxes or variations in line ratio values than $n_{\rm{H}_2}$, which may imply that the information provided from two low-J CO lines may be insufficient to well-constraint $T_{\rm{K}}$ and that a larger, multiline study, like that of \citet{Zhang14} would likely substantially improve $T_{\rm{K}}$ estimations. In addition, the use of only two low excitation CO transitions in our RADEX modeling could be responsible for the relatively low ($T_{\rm{K}} \lesssim 10$~K) temperatures seen in the spiral arms and inter-arm regions as well as a lower-than-expected nuclear temperature \citep[e.g.,][]{Tan11, Zhang14}.

\begin{figure*}[ht]
\centering
\includegraphics[scale=0.6]{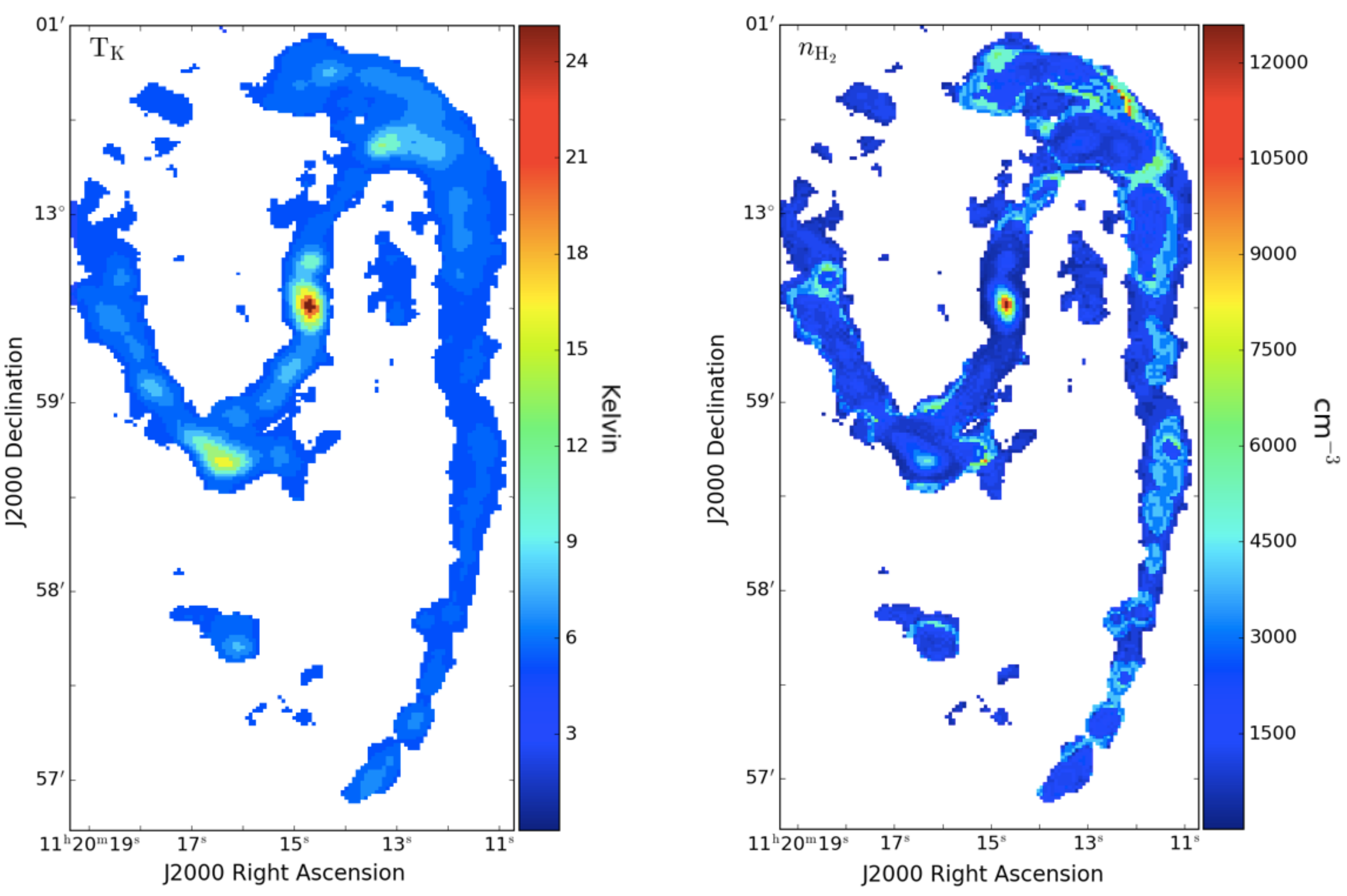}
\caption{\textit{Left}: Kinetic temperature map of NGC 3627, derived from RADEX modeling. \textit{Right}: H$_2$ number density map of NGC 3627, derived from RADEX modeling. Each pixel in both maps is ${\sim}1^{\prime \prime} \times 1^{\prime \prime}$, which corresponds to an approximate physical scale of ${\sim}51\,\rm{pc} \times 51 \rm{pc}$.} 
\label{fig:RADEX_maps}
\end{figure*}

\subsubsection{Correlations between SFE and Physical Parameters}

\begin{figure*}[ht]
\centering
\includegraphics[scale=0.65]{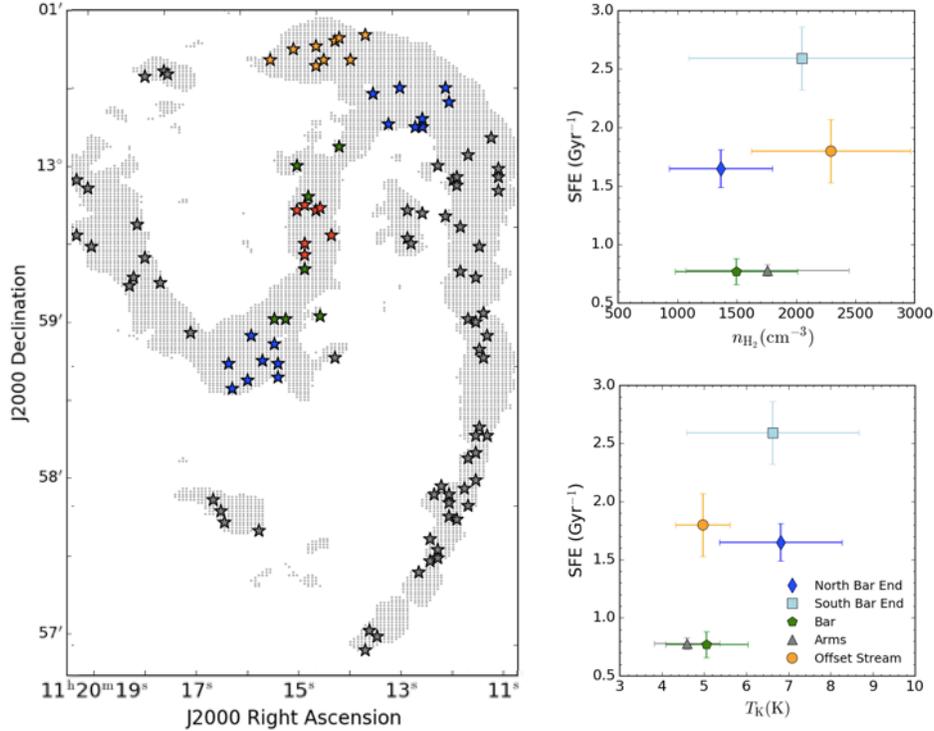}
\caption{\textit{Left}: A representative MC trial for each of the six regions of interest in NGC~3627. The colors, regions, and number of pixel samples per region have been chosen to match those used in \citet{Watanabe11}. \textit{Right}: Correlations between SFE and $T_{\rm{K}}$ (bottom) and $n_{\rm{H}_2}$ (top). If the central data point is included, no significant correlation is observed, but if the central data point is excluded as is shown, $T_{\rm{K}}$ is found to be tentatively correlated with SFE.}
\label{fig:SFE_plot}
\end{figure*}

We examine the dependence of star formation efficiency (SFE) on the physical parameters $T_{\rm{K}}$ and $n_{\rm{H}_2}$ by investigating potential correlations across different spatial regions in NGC~3627. We take the SFE values from \citet{Watanabe11}, who derive typical values for six different regions in NGC~3627 using integrated $J=1-0$ line intensities of the $^{12}$CO and $^{13}$CO transitions. As shown in Figure \ref{fig:RADEX_maps}, there are order-of-magnitude spatial variations in $T_{\rm{K}}$ and $n_{\rm{H}_2}$, which makes estimating typical values difficult. We separated our RADEX-derived maps of NGC~3627 into six representative regions considered by \citet{Watanabe11}, namely the center, bar, southern bar end, northern bar end, offset stream, and spiral arms. Then, we randomly sampled each region and recorded the values of $T_{\rm{K}}$ and $n_{\rm{H}_2}$. To ensure consistency with \citet{Watanabe11}, we chose to sample the same number of pixels per region (see their Figure 4). Adopting a Monte Carlo (MC) approach, we repeated this random sampling for $10^4$ trials for each region. To estimate typical values and uncertainties for each region, we then computed the mean and standard deviation of all MC runs. The left panel of Figure \ref{fig:SFE_plot} shows a representative MC trial with each color corresponding to a different sampling region.

We calculate Pearson's correlation coefficients, which characterize linear relationships, to evaluate the significance of the correlations. We do not find correlations between either SFE-$T_{\rm{K}}$ or SFE-$n_{\rm{H}_2}$ with coefficients of $r=0.25$ and $r=0.15$, respectively. However, the central data point is a substantial outlier both in terms of $T_{\rm{K}}$ and $n_{\rm{H}_2}$ and has the largest uncertainties $(>50\%)$. This nuclear uncertainty is explained, especially in the case of $n_{\rm{H}_2}$, by the abrupt and order-of-magnitude increase in density in the innermost ${\sim}$150~pc region of NGC~3627. In our MC trials, we are occasionally selecting values from this inner region (i.e., $n_{\rm{H}_2} >10000$ cm$^{-3}$; $T_{\rm{K}} > 20~\rm{K}$), which inflates standard deviations over the course of the MC trials.
 
If we exclude this outlying central data point, we find correlation coefficients of $r=0.79$ and $r=0.29$ for $T_{\rm{K}}$ and $n_{\rm{H}_2}$, respectively. The former correlation between SFE and $T_{\rm{K}}$ is significant at the ${\sim}$90\% confidence level. If instead, we use the MC medians, which are typically 15--20\% lower than the means, to represent characteristic $T_{\rm{K}}$ and $n_{\rm{H}_2}$ values when analyzing correlations, the $r$ values remain approximately the same and the significance of the SFE-$n_{\rm{H}_2}$ correlation is unchanged. If we instead chose the peak values of $T_{\rm{K}}$ and $n_{\rm{H}_2}$ in each region, the SFE-$T_{\rm{K}}$ correlation becomes more tentative ($r=0.65$) and the SFE-$n_{\rm{H}_2}$ remains weak ($r=0.50$).

Thus, as long as the nuclear value is excluded, the tentative correlation between $T_{\rm{K}}$ and SFE is robust across several proxies for physical parameters in NGC~3627. Interestingly, we did not find a correlation between SFE and $n_{\rm{H}_2}$, such as the strong one seen in NGC~2903 \citep{Muraoka16}. Such a correlation may have been expected, since molecular gas density is thought to control spatial variations in SFE, based on numerous previous studies with HCN \citep[e.g.,][]{Gao04, Gao07, Muraoka09, Usero15} and CO \citep[e.g.,][]{Muraoka16, Koyama17}. However, being an interacting galaxy, NGC~3627 may follow a more complex scaling relationship between $n_{\rm{H}_2}$ and SFE. The reported correlations (and lack thereof) should be treated as somewhat speculative in nature and while interesting in their own right, deserve additional follow-up in NGC~3627 and across a larger set of galaxies with observations of large-scale, multiline $^{12}$CO emission.
 
\subsubsection{Molecular Gas Estimates}

By using $^{12}$CO($J=1-0$) observations from the IRAM 30~m telescope along with an H$_2$--CO conversion factor of $2.2\times10^{20}$ cm$^{-2}$, \citet{Casasola11} derived an H$_2$ mass of $9.9\times10^{8}M_{\odot}$ within the inner ${\sim}50^{\prime \prime} \times 50^{\prime \prime}$ of the galaxy. To compare our modeling results, we sum all $n_{\rm{H}_2}$ pixels in Figure \ref{fig:RADEX_maps} within the same ${\sim}50^{\prime \prime} \times 50^{\prime \prime}$ region and normalize by the beamsize. We assume a column length of 440~pc (${\sim}8.6^{\prime \prime}$), which corresponds to the vertical FWHM of CO gas in edge-on spiral galaxy NGC~891 \citep{Yim11}. We find $M_{\rm{H}_2} \sim 5.6 \times 10^9 M_{\odot}$, which is only ${\sim}40$\% lower than the value derived by \citet{Casasola11}.

We can also estimate the total galactic H$_2$ mass using the same method and find a total H$_2$ mass estimate of $9.6 \times 10^9 M_{\odot}$. As we do not know the true vertical distribution of molecular gas in NGC~3627, the greatest uncertainty is related to the assumed column length. An uncertainty of $\pm1^{\prime \prime}$ in column length leads to a ${\sim}$10--15\% uncertainty in total H$_2$ mass. However, as we do not have detailed information about the three-dimensional structure of NGC 3627, this method represents the most reasonable estimate of total H$_2$ mass that we can make. If we compare our mass determination to previous estimates of $4.1 \pm 0.41 \times 10^9 M_{\odot}$ \citep{Helfer03} and $4.9 \times 10^9 M_{\odot}$ \citep{Kuno07}, we find that our estimates are about 50\% larger. While we choose a representative CO FWHM from \citet{Yim14} (see their Figure 14) from NGC~891 since it has a similar H$_2$ mass as NGC~3627 \citep[${\sim}3.8\times10^9 M_{\odot}$;][]{Scoville93}, if we had instead chosen a FWHM of ${\sim}$250~pc as for NGC~5907 or ${\sim}110$~pc as for NGC~4157 and NGC~4565, derived H$_2$ mass estimates are ${\sim}40\%$ and ${\sim}75\%$ lower, respectively.

We can also estimate the mass of the isolated clump of $^{12}$CO$(J=2-1)$ emission. This isolated \ion{H}{2} region has been observed in the $J=1-0$ transitions of HCN and HCO$^+$ and is converting its molecular gas into stars more efficiently by a factor of ${\sim}$3 than either the star-forming regions in the southern bar end or nuclear disk \citep{Murphy15}. There are also likely large reserves of ultra-dense gas surrounding this newly-formed \ion{H}{2} region \citep{Murphy15}, which makes it crucial to constrain the molecular mass in this region. Assuming the same vertical scale as above, we derive an H$_2$ mass of $2.4 \times 10^8 M_{\odot}$.

Due to the abundance of independent H$_2$ mass estimates of NGC~3627, we can also constrain the vertical height of its constituent molecular gas. To recover the $M_{\rm{H}_2} = 9.9\times10^{8}M_{\odot}$ that \citet{Casasola11} estimated in the inner region of NGC~3627, we require an unusually large vertical height of ${\sim}770$~pc, which implies that our RADEX calculations may be underestimating $n_{\rm{H_2}}$ in the high-density, high-temperature nuclear region. Alternatively, the nuclear region of NGC~3627 may not be well-described by a constant H$_2$--CO conversion factor, leading to differences in mass determinations. Similarly, the H$_2$ mass estimate of $4.9\times10^9M_{\odot}$, which was based on CO data from the Nobeyama 45~m telescope \citep{Kuno07}, would require a uniform vertical extent in CO of ${\sim}$225~pc over the whole galaxy. While this is an order-of-magnitude estimation, especially since we are ignoring the complexities often seen in vertical molecular gas distributions, i.e. increasing scale height with galactocentric radius \citep[e.g.,][]{Scoville93}, this finding is consistent with the ${\sim}100$--$450$~pc range in molecular gas FWHMs derived by \citet{Yim14}.

\subsection{NGC 3627 Rotation Curve}
\label{sec:rotation_curve_section}
We ran the kinemetry IDL routine \citep{Krajnovic06} on the \textit{uv}-tapered, moment-1 map of NGC~3627 to estimate a galactic rotation curve. The program performs a harmonic expansion of 2D moment maps along best-fitting ellipses to quantify line-of-sight velocity distributions and identify morphological and kinematic components. The inclination of each ellipse is based on the axial ratio $q$ via $\cos i = q$ with $q=1$ (a circle) indicating a face-on inclination $i=0^{\circ}$, while $q=0$ (a line) yields an edge-on inclination of $i=90^{\circ}$. Thus, under this definition, the inclination of NGC~3627 would correspond to an $i$ of $90^{\circ} - 61.3^{\circ} = 28.7^{\circ}$ and $q \approx 0.877$.

We ran two iterations of kinemetry, one in which we fixed the inclination of all best-fitting ellipses to $q=0.877$ and another in which the inclination was a free parameter and allowed to vary for each ellipse. We allowed the position angles of the ellipses to vary for both iterations. We also restricted the radii of the best-fitting ellipses to only sample distances approximately equal to one-half of the smallest dimension of the beam size, i.e. $\frac{1}{2} \times 5.22^{\prime \prime} = 2.61^{\prime \prime}$. In order to fit larger galactocentric radii, we also relaxed the default `cover' parameter from 0.75 to 0.4, which means that if less than 40\% of the points along an ellipse are not present, the program will stop. We find systemic velocities of $717.85 \pm 20.36$ km s$^{-1}$ and $721.52 \pm 20.32$ km s$^{-1}$ as well as kinematic PAs of $172.48^{\circ} \pm 9.30^{\circ}$ and $177.84^{\circ} \pm 8.02^{\circ}$ for the fixed and free-to-vary $i$ kinemetry runs. The derived values and uncertainties are given as the mean and standard deviation, respectively, of the fitted ellipses. Both values are consistent with those derived by \citet{Casasola11}, although the freely-varying inclination iteration more closely recovers the expected $\rm{PA}=178^{\circ}$ of NGC~3627.

The systemic velocities derived from the $^{12}$CO$(J=2-1)$ emission are more consistent with $v_{\rm{hel}}=$720 km s$^{-1}$ obtained from the \ion{H}{1} data from \citet{Haan08}. \citet{Casasola11} report a systemic velocity that is redshifted by 24 km s$^{-1}$ from 720 km s$^{-1}$ and note that interacting galaxies and those with asymmetric \ion{H}{1} morphology often exhibit discrepant $^{12}$CO and \ion{H}{1} systemic velocities, such as the 50 km s$^{-1}$ differences seen in NGC~4579 and NGC~5953 by \citet{Casasola10} and \citet{Garcia09}, respectively. They attribute these differences to interaction history and the differing sensitivities of atomic and molecular gas to ram pressure effects \citep[and references therein]{Garcia09}. Despite our ${\sim}$20 km~s$^{-1}$ uncertainties on $v_{\rm{hel}}$, the fact that our systemic velocity determinations are more consistent with those derived from \ion{H}{1} implies that either ram-pressure effects may be overstated by \citet{Casasola11} or there are differences in bulk velocity between the $J=2-1$ and $J=1-0$ transitions of $^{12}$CO. Since this latter effect was also seen in NGC~4579 by \citet{Garcia09}, who reported that the $J=1-0$ velocity centriod was blueshifted by 20 km~s$^{-1}$ from that of $J=2-1$, it is difficult to definitively comment on the reasons for differences in systemic velocity determinations.

\begin{figure}[htp!]
\centering
\includegraphics[scale=0.45]{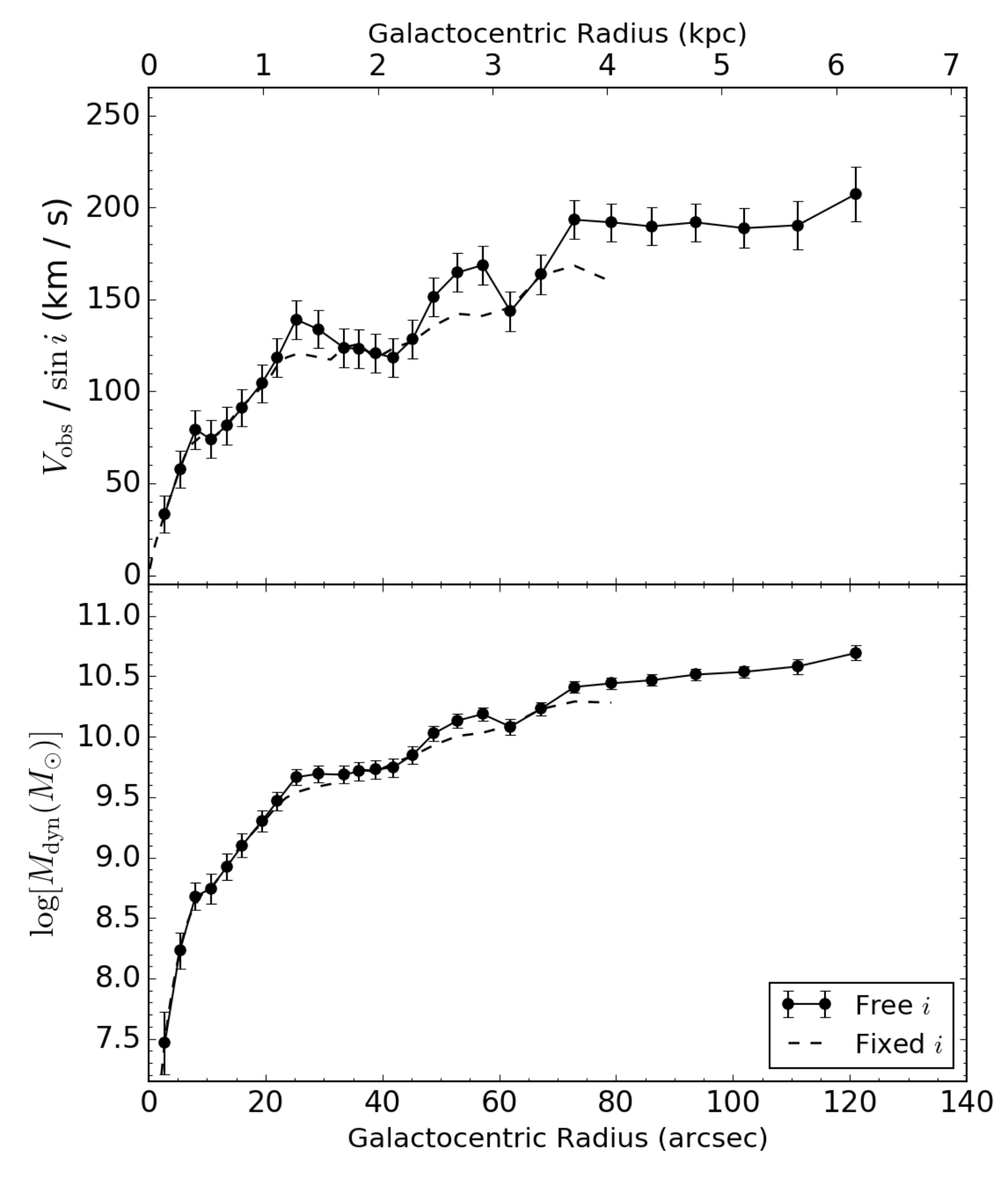}
\caption{\textit{Top}: Deprojected rotation curves for NGC 3627, derived with kinemetry. \textit{Bottom}: Dynamical mass estimates for each galactocentric radius. In some cases, the error bars are smaller than the marker size. For both panels, rotation curves and mass estimates were derived with best-fitting ellipse inclination allowed to vary (solid line) and with inclination fixed to that of NGC~3627 (dashed line).}
\label{fig:Whole_Galaxy_Rot}
\end{figure}

The top panel of Figure \ref{fig:Whole_Galaxy_Rot} shows the derived rotation curve with inclination-corrected velocities ($V_{\rm{obs}} / \sin i$). While both kinemetry runs estimate nearly the same velocities within 1~kpc, the fixed $i$ iteration finds velocities that are ${\sim}20$--$30$ km s$^{-1}$ lower than the freely-varying $i$ run. Otherwise, the behavior of both rotation curves is essentially identical. The fixed $i$ rotation curve also only probes to a galactocentric radius of 4 kpc, while the freely varying $i$ iteration extends to a little past 6~kpc. The free-to-vary $i$ run probes ${\sim}35\%$ further than the fixed $i$ run, because the script is permitted to alter the axial flattening ratio $q$ (i.e., the ellipse shape) for each best-fitting ellipse. Whenever the ellipses reach the edge of the moment-1 map, kinemetry can change their shapes to encompass more velocity values such that it is still consistent with the data and is able to fit additional higher-radii ellipses. However, the fixed $i$ run adopts a constant ellipse shape, which cannot be changed even as the ellipses reach the map edge, and thus, more quickly terminates as the cover parameter is exceeded.

For both runs, the signature of a rigidly rotating nuclear bar is seen in the first ${\sim}20^{\prime \prime}$ as well as several dips which likely correspond to non-circular motions at the transitional zones between the nucleus, inter-arm regions, bar ends, and spirals. For the freely varying $i$ run, a velocity of ${\sim}$190 km s$^{-1}$ is achieved at ${\sim}3.5$~kpc. After this point, the rotation curve levels off, increasing by a few ${\sim}$tens of km s$^{-1}$ at $\gtrsim 6$~kpc, and reaching a peak of ${\sim}207$~km~s$^{-1}$ at ${\sim}6.2$~kpc.

We can estimate the dynamical mass within various radii via the formula $M(R) = 2.325\times10^5 \alpha R V^2(R)$, where $M(R)$ is the mass in $M_{\odot}$ within radius $R$, $R$ is in kpc, and $V$ is in km s$^{-1}$, and $\alpha$ is a factor which encodes the geometry of the system. We adopt $\alpha=0.8$, an intermediate value between a spherical distribution (1.0) and a flat dist (0.6), for consistency with \citet{Casasola11}. The bottom panel of Figure \ref{fig:Whole_Galaxy_Rot} shows dynamical mass at each galactocentric radius. Within $21^{\prime \prime}$, \citet{Casasola11} found a dynamical mass estimate of $6.0 \times 10^9 M_{\odot}$, while we report an average $M_{\rm{dyn}} = 2.83 \pm 0.51 \times 10^9 M_{\odot}$ at $22^{\prime \prime}$ for the fixed and free inclination runs, respectively. However, \citet{Casasola11} assumed an extrapolated maximum velocity of 180 km s$^{-1}$ to estimate this value, while we have used a velocity of 114 km s$^{-1}$ (fixed $i$) and 118 km s$^{-1}$ (free $i$) at $22^{\prime \prime}$ as determined by kinemetry. At ${\sim}$6.2~kpc, the furthest distance probed by the free $i$ run, we find a total dynamical mass of $4.94 \pm 0.70 \times 10^{10} M_{\odot}$.

To estimate the entire dynamical mass of the galaxy, we assume a flat rotation curve at large radii, which is consistent with the flattening behavior we see out to a galactocentric radius of 6.2 kpc. We adopt the median velocity of ${\sim}192$ km s$^{-1}$ for all velocity estimates past 3.5 kpc, as this is when the rotation curve beings to flatten, as our maximum velocity. The furthest emission that we detect lies at ${\sim}165^{\prime \prime}$, which is about 8.4 kpc from the galactic center. As we do not actually observe the rotational velocities at this distance, we report an upper limit to the total dynamical mass for NGC 3627 of $5.75 \times 10^{10} M_{\odot}$.

\section{Conclusions}
\label{sec:conclusions}
We presented a $^{12}$CO($J=2-1$) emission map for NGC 3627. Based on CO observations of the interacting spiral galaxy NGC 3627 with the SMA, we conclude the following: 

\begin{itemize}

\item We find enhanced emission and velocity dispersions in the nuclear region and bar ends with more diffuse emission and smaller dispersion in the spiral arms. We find a velocity gradient of ${\sim}400$--$450$ km s$^{-1}$ across the entire galaxy.

\item We detected unresolved $^{13}$CO($J=2-1$) emission in the galactic center, southern bar end, and in an isolated clump of emission in the south of NGC 3627. Typical integrated-line intensity ratios of $^{12}$CO / $^{13}$CO are ${\sim}2.5$--$4$ with elevated ratios corresponding to regions with higher $\Sigma_{\rm{SFR}}$. No C$^{18}$O($J=2-1$) emission was detected in NGC~3627 down to a $3\sigma$ rms noise level of 42 mJy beam$^{-1}$ per 20~km~s$^{-1}$ channel.

\item Using archival BIMA $^{12}$CO($J=1-0$), we produced a $R_{21/10}$ line ratio map for NGC~3627. Moderate ratio gas (${\sim}$0.4--0.6) was found in the nuclear region and the bar ends, while the spiral and inter-arm regions often exhibited substantially lower ratios ${\sim}$0.2. High ratio gas was also observed in the southernmost end of the extended western arm (${\gtrsim}$\,1.0) and in the clump (${\sim}$0.6--0.7), indicative of warm and dense molecular material likely due to previous tidal interaction with NGC~3628.

\item Using the $J=2-1$ and $J=1-0$ transitions of $^{12}$CO, we produced a map of beam-averaged kinetic temperature and $n_{\rm{H}_2}$ at physical scales of ${\sim}$50 pc under non-LTE conditions. Kinetic temperatures ranged from ${\sim}5$--$10$ K in the spiral arms to ${\sim}25$ K in the nuclear region. A similar trend was found for $n_{\rm{H}_2}$ with values spanning ${\sim}$400--1000 cm$^{-3}$ to ${\sim}$12500 cm$^{-3}$.

\item For all regions except the center, we find a tentative SFE-$\rm{T}_{\rm{K}}$ correlation and no correlation between SFE and $n_{\rm{H}_2}$. Since molecular gas density is believed to control spatial variations observed in SFE, the lack of this latter correlation is potentially surprising but with the important caveat that our RADEX analysis based on two $^{12}$CO lines, i.e. $J=2-1$ and $J=1-0$, may allow us to only probe a narrow range of gas volume densities.

\item We derived a rotation curve for NGC 3627 out to a galactocentric radius of ${\sim}6.2$ kpc. Assuming an intermediate geometry between a flat disk and spherical distribution, we estimated dynamical mass as a function of galactocentric radius. By using the median velocity of ${\sim}192$ km s$^{-1}$ of the flat portion of the rotation curve at large galactocentric radius ($>3.5$ kpc), we report an upper limit of $M_{\rm{dyn}} = 5.75 \times 10^{10} M_{\odot}$ for the entire galaxy.

\end{itemize}

\acknowledgments

We thank the anonymous referee for the helpful comments that improved the content and presentation of this work. We thank the observers who were involved in acquiring the SMA observations. C.J.L. thanks Davor Krajnovi\'c for helpful discussions concerning the use of his kinemetry IDL program and Hauyu Baobab Liu for his insightful comments regarding the feathering of single-dish and interferometric data. This work made use of HERACLES, `The HERA CO-Line Extragalactic Survey' \citep{Leroy09}. This research has made use of the NASA/IPAC Extragalactic Database (NED), which is operated by the Jet Propulsion Laboratory, California Institute of Technology, under contract with the National Aeronautics and Space Administration.

%

\vspace{5mm}
\facilities{SMA}


\software{MIR (\url{http:
//www.cfa.harvard.edu/~cqi/mircook.html}), CASA \citep{McMullin07}, myRADEX (\url{https://github.com/fjdu/myRadex}), kinemetry \citep{Krajnovic06}}



\bibliography{SMAbib}

\begin{thebibliography}{}
\expandafter\ifx\csname natexlab\endcsname\relax\def\natexlab#1{#1}\fi
\providecommand{\url}[1]{\href{#1}{#1}}

\bibitem[{{Arp}(1966)}]{Arp66}
{Arp}, H. 1966, \apjs, 14, 1

\bibitem[{{Beuther} {et~al.}(2017){Beuther}, {Meidt}, {Schinnerer}, {Paladino},
  \& {Leroy}}]{Beuther17}
{Beuther}, H., {Meidt}, S., {Schinnerer}, E., {Paladino}, R., \& {Leroy}, A.
  2017, \aap, 597, A85

\bibitem[{{Blake} {et~al.}(1987){Blake}, {Sutton}, {Masson}, \&
  {Phillips}}]{Blake87}
{Blake}, G.~A., {Sutton}, E.~C., {Masson}, C.~R., \& {Phillips}, T.~G. 1987,
  \apj, 315, 621

\bibitem[{{Bridge} {et~al.}(2010){Bridge}, {Carlberg}, \&
  {Sullivan}}]{Bridge10}
{Bridge}, C.~R., {Carlberg}, R.~G., \& {Sullivan}, M. 2010, \apj, 709, 1067

\bibitem[{{Calzetti} {et~al.}(2015){Calzetti}, {Lee}, {Sabbi}, {Adamo},
  {Smith}, {Andrews}, {Ubeda}, {Bright}, {Thilker}, {Aloisi}, {Brown},
  {Chandar}, {Christian}, {Cignoni}, {Clayton}, {da Silva}, {de Mink}, {Dobbs},
  {Elmegreen}, {Elmegreen}, {Evans}, {Fumagalli}, {Gallagher}, {Gouliermis},
  {Grebel}, {Herrero}, {Hunter}, {Johnson}, {Kennicutt}, {Kim}, {Krumholz},
  {Lennon}, {Levay}, {Martin}, {Nair}, {Nota}, {{\"O}stlin}, {Pellerin},
  {Prieto}, {Regan}, {Ryon}, {Schaerer}, {Schiminovich}, {Tosi}, {Van Dyk},
  {Walterbos}, {Whitmore}, \& {Wofford}}]{Calzetti15}
{Calzetti}, D., {Lee}, J.~C., {Sabbi}, E., {et~al.} 2015, \aj, 149, 51

\bibitem[{{Casasola} {et~al.}(2004){Casasola}, {Bettoni}, \&
  {Galletta}}]{Casasola04}
{Casasola}, V., {Bettoni}, D., \& {Galletta}, G. 2004, \aap, 422, 941

\bibitem[{{Casasola} {et~al.}(2010){Casasola}, {Hunt}, {Combes},
  {Garc{\'{\i}}a-Burillo}, {Boone}, {Eckart}, {Neri}, \&
  {Schinnerer}}]{Casasola10}
{Casasola}, V., {Hunt}, L.~K., {Combes}, F., {et~al.} 2010, \aap, 510, A52

\bibitem[{{Casasola} {et~al.}(2011){Casasola}, {Hunt}, {Combes},
  {Garc{\'{\i}}a-Burillo}, \& {Neri}}]{Casasola11}
{Casasola}, V., {Hunt}, L.~K., {Combes}, F., {Garc{\'{\i}}a-Burillo}, S., \&
  {Neri}, R. 2011, \aap, 527, A92

\bibitem[{{Chemin} {et~al.}(2003){Chemin}, {Cayatte}, {Balkowski}, {Marcelin},
  {Amram}, {van Driel}, \& {Flores}}]{Chemin03}
{Chemin}, L., {Cayatte}, V., {Balkowski}, C., {et~al.} 2003, \aap, 405, 89

\bibitem[{{Cormier} {et~al.}(2018){Cormier}, {Bigiel}, {Jim{\'e}nez-Donaire},
  {Leroy}, {Gallagher}, {Usero}, {Sandstrom}, {Bolatto}, {Hughes}, {Kramer},
  {Krumholz}, {Meier}, {Murphy}, {Pety}, {Rosolowsky}, {Schinnerer}, {Schruba},
  {Sliwa}, \& {Walter}}]{Cormier18}
{Cormier}, D., {Bigiel}, F., {Jim{\'e}nez-Donaire}, M.~J., {et~al.} 2018,
  \mnras, 475, 3909

\bibitem[{{Dahlem} {et~al.}(1996){Dahlem}, {Heckman}, {Fabbiano}, {Lehnert}, \&
  {Gilmore}}]{Dahlem96}
{Dahlem}, M., {Heckman}, T.~M., {Fabbiano}, G., {Lehnert}, M.~D., \& {Gilmore},
  D. 1996, \apj, 461, 724

\bibitem[{{Davis}(2014)}]{Davis14}
{Davis}, T.~A. 2014, \mnras, 445, 2378

\bibitem[{{de Vaucouleurs} {et~al.}(1991){de Vaucouleurs}, {de Vaucouleurs},
  {Corwin}, {Buta}, {Paturel}, \& {Fouqu{\'e}}}]{Vaucouleurs91}
{de Vaucouleurs}, G., {de Vaucouleurs}, A., {Corwin}, Jr., H.~G., {et~al.}
  1991, {Third Reference Catalogue of Bright Galaxies. Volume I: Explanations
  and references. Volume II: Data for galaxies between 0$^{h}$ and 12$^{h}$.
  Volume III: Data for galaxies between 12$^{h}$ and 24$^{h}$.}

\bibitem[{{Dumke} {et~al.}(2011){Dumke}, {Krause}, {Beck}, {Soida}, {Urbanik},
  \& {Wielebinski}}]{Dumke11}
{Dumke}, M., {Krause}, M., {Beck}, R., {et~al.} 2011, in Astronomical Society
  of the Pacific Conference Series, Vol. 446, Galaxy Evolution: Infrared to
  Millimeter Wavelength Perspective, ed. W.~{Wang}, J.~{Lu}, Z.~{Luo},
  Z.~{Yang}, H.~{Hua}, \& Z.~{Chen}, 111

\bibitem[{{Elmegreen} \& {Elmegreen}(1987)}]{Elmegreen87}
{Elmegreen}, D.~M., \& {Elmegreen}, B.~G. 1987, \apj, 314, 3

\bibitem[{{Filippenko} \& {Sargent}(1985)}]{Filippenko85}
{Filippenko}, A.~V., \& {Sargent}, W.~L.~W. 1985, \apjs, 57, 503

\bibitem[{{Gallagher} {et~al.}(2018){Gallagher}, {Leroy}, {Bigiel}, {Cormier},
  {Jim{\'e}nez-Donaire}, {Ostriker}, {Usero}, {Bolatto},
  {Garc{\'{\i}}a-Burillo}, {Hughes}, {Kepley}, {Krumholz}, {Meidt}, {Meier},
  {Murphy}, {Pety}, {Rosolowsky}, {Schinnerer}, {Schruba}, \&
  {Walter}}]{Gallagher18}
{Gallagher}, M.~J., {Leroy}, A.~K., {Bigiel}, F., {et~al.} 2018, \apj, 858, 90

\bibitem[{{Gao} {et~al.}(2007){Gao}, {Carilli}, {Solomon}, \& {Vanden
  Bout}}]{Gao07}
{Gao}, Y., {Carilli}, C.~L., {Solomon}, P.~M., \& {Vanden Bout}, P.~A. 2007,
  \apjl, 660, L93

\bibitem[{{Gao} \& {Solomon}(2004)}]{Gao04}
{Gao}, Y., \& {Solomon}, P.~M. 2004, \apjs, 152, 63

\bibitem[{{Garc{\'{\i}}a-Burillo} {et~al.}(2009){Garc{\'{\i}}a-Burillo},
  {Fern{\'a}ndez-Garc{\'{\i}}a}, {Combes}, {Hunt}, {Haan}, {Schinnerer},
  {Boone}, {Krips}, \& {M{\'a}rquez}}]{Garcia09}
{Garc{\'{\i}}a-Burillo}, S., {Fern{\'a}ndez-Garc{\'{\i}}a}, S., {Combes}, F.,
  {et~al.} 2009, \aap, 496, 85

\bibitem[{{Georgantopoulos} {et~al.}(2002){Georgantopoulos}, {Panessa},
  {Akylas}, {Zezas}, {Cappi}, \& {Comastri}}]{Georgantopoulos02}
{Georgantopoulos}, I., {Panessa}, F., {Akylas}, A., {et~al.} 2002, \aap, 386,
  60

\bibitem[{{Gil de Paz} {et~al.}(2007){Gil de Paz}, {Boissier}, {Madore},
  {Seibert}, {Joe}, {Boselli}, {Wyder}, {Thilker}, {Bianchi}, {Rey}, {Rich},
  {Barlow}, {Conrow}, {Forster}, {Friedman}, {Martin}, {Morrissey}, {Neff},
  {Schiminovich}, {Small}, {Donas}, {Heckman}, {Lee}, {Milliard}, {Szalay}, \&
  {Yi}}]{Gil07}
{Gil de Paz}, A., {Boissier}, S., {Madore}, B.~F., {et~al.} 2007, \apjs, 173,
  185

\bibitem[{{Goldreich} \& {Kwan}(1974)}]{Goldreich74}
{Goldreich}, P., \& {Kwan}, J. 1974, \apj, 189, 441

\bibitem[{{Haan} {et~al.}(2009){Haan}, {Schinnerer}, {Emsellem},
  {Garc{\'{\i}}a-Burillo}, {Combes}, {Mundell}, \& {Rix}}]{Haan09}
{Haan}, S., {Schinnerer}, E., {Emsellem}, E., {et~al.} 2009, \apj, 692, 1623

\bibitem[{{Haan} {et~al.}(2008){Haan}, {Schinnerer}, {Mundell},
  {Garc{\'{\i}}a-Burillo}, \& {Combes}}]{Haan08}
{Haan}, S., {Schinnerer}, E., {Mundell}, C.~G., {Garc{\'{\i}}a-Burillo}, S., \&
  {Combes}, F. 2008, \aj, 135, 232

\bibitem[{{Haynes} {et~al.}(1979){Haynes}, {Giovanelli}, \&
  {Roberts}}]{Haynes79}
{Haynes}, M.~P., {Giovanelli}, R., \& {Roberts}, M.~S. 1979, \apj, 229, 83

\bibitem[{{Helfer} {et~al.}(2003){Helfer}, {Thornley}, {Regan}, {Wong},
  {Sheth}, {Vogel}, {Blitz}, \& {Bock}}]{Helfer03}
{Helfer}, T.~T., {Thornley}, M.~D., {Regan}, M.~W., {et~al.} 2003, \apjs, 145,
  259

\bibitem[{{Ho} {et~al.}(1997){Ho}, {Filippenko}, \& {Sargent}}]{Ho97}
{Ho}, L.~C., {Filippenko}, A.~V., \& {Sargent}, W.~L.~W. 1997, \apjs, 112, 315

\bibitem[{{Ho} {et~al.}(2004){Ho}, {Moran}, \& {Lo}}]{Ho04}
{Ho}, P.~T.~P., {Moran}, J.~M., \& {Lo}, K.~Y. 2004, \apjl, 616, L1

\bibitem[{{Hunt} {et~al.}(2017){Hunt}, {Wei{\ss}}, {Henkel}, {Combes},
  {Garc{\'{\i}}a-Burillo}, {Casasola}, {Caselli}, {Lundgren}, {Maiolino},
  {Menten}, \& {Testi}}]{Hunt17}
{Hunt}, L.~K., {Wei{\ss}}, A., {Henkel}, C., {et~al.} 2017, \aap, 606, A99

\bibitem[{{Israel}(2009{\natexlab{a}})}]{Israel09a}
{Israel}, F.~P. 2009{\natexlab{a}}, \aap, 506, 689

\bibitem[{{Israel}(2009{\natexlab{b}})}]{Israel09b}
---. 2009{\natexlab{b}}, \aap, 493, 525

\bibitem[{{Jim{\'e}nez-Donaire}
  {et~al.}(2017{\natexlab{a}}){Jim{\'e}nez-Donaire}, {Cormier}, {Bigiel},
  {Leroy}, {Gallagher}, {Krumholz}, {Usero}, {Hughes}, {Kramer}, {Meier},
  {Murphy}, {Pety}, {Schinnerer}, {Schruba}, {Schuster}, {Sliwa}, \&
  {Tomicic}}]{Donaire17}
{Jim{\'e}nez-Donaire}, M.~J., {Cormier}, D., {Bigiel}, F., {et~al.}
  2017{\natexlab{a}}, \apjl, 836, L29

\bibitem[{{Jim{\'e}nez-Donaire}
  {et~al.}(2017{\natexlab{b}}){Jim{\'e}nez-Donaire}, {Bigiel}, {Leroy},
  {Cormier}, {Gallagher}, {Usero}, {Bolatto}, {Colombo},
  {Garc{\'{\i}}a-Burillo}, {Hughes}, {Kramer}, {Krumholz}, {Meier}, {Murphy},
  {Pety}, {Rosolowsky}, {Schinnerer}, {Schruba}, {Tomi{\v c}i{\'c}}, \&
  {Zschaechner}}]{2017MNRAS.466...49J}
{Jim{\'e}nez-Donaire}, M.~J., {Bigiel}, F., {Leroy}, A.~K., {et~al.}
  2017{\natexlab{b}}, \mnras, 466, 49

\bibitem[{{Karachentsev} \& {Kudrya}(2014)}]{Karachentsev14}
{Karachentsev}, I.~D., \& {Kudrya}, Y.~N. 2014, \aj, 148, 50

\bibitem[{{Kennicutt} {et~al.}(2003){Kennicutt}, {Armus}, {Bendo}, {Calzetti},
  {Dale}, {Draine}, {Engelbracht}, {Gordon}, {Grauer}, {Helou}, {Hollenbach},
  {Jarrett}, {Kewley}, {Leitherer}, {Li}, {Malhotra}, {Regan}, {Rieke},
  {Rieke}, {Roussel}, {Smith}, {Thornley}, \& {Walter}}]{Kennicutt03}
{Kennicutt}, Jr., R.~C., {Armus}, L., {Bendo}, G., {et~al.} 2003, \pasp, 115,
  928

\bibitem[{{Koyama} {et~al.}(2017){Koyama}, {Koyama}, {Yamashita},
  {Morokuma-Matsui}, {Matsuhara}, {Nakagawa}, {Hayashi}, {Kodama}, {Shimakawa},
  {Suzuki}, {Tadaki}, {Tanaka}, \& {Yamamoto}}]{Koyama17}
{Koyama}, S., {Koyama}, Y., {Yamashita}, T., {et~al.} 2017, \apj, 847, 137

\bibitem[{{Krajnovi{\'c}} {et~al.}(2006){Krajnovi{\'c}}, {Cappellari}, {de
  Zeeuw}, \& {Copin}}]{Krajnovic06}
{Krajnovi{\'c}}, D., {Cappellari}, M., {de Zeeuw}, P.~T., \& {Copin}, Y. 2006,
  \mnras, 366, 787

\bibitem[{{Krips} {et~al.}(2008){Krips}, {Neri}, {Garc{\'{\i}}a-Burillo},
  {Mart{\'{\i}}n}, {Combes}, {Graci{\'a}-Carpio}, \& {Eckart}}]{Krips08}
{Krips}, M., {Neri}, R., {Garc{\'{\i}}a-Burillo}, S., {et~al.} 2008, \apj, 677,
  262

\bibitem[{{Kuno} {et~al.}(2007){Kuno}, {Sato}, {Nakanishi}, {Hirota}, {Tosaki},
  {Shioya}, {Sorai}, {Nakai}, {Nishiyama}, \& {Vila-Vilar{\'o}}}]{Kuno07}
{Kuno}, N., {Sato}, N., {Nakanishi}, H., {et~al.} 2007, \pasj, 59, 117

\bibitem[{{Lee} \& {Jang}(2013)}]{Lee13}
{Lee}, M.~G., \& {Jang}, I.~S. 2013, \apj, 773, 13

\bibitem[{{Leroy} {et~al.}(2009){Leroy}, {Walter}, {Bigiel}, {Usero}, {Weiss},
  {Brinks}, {de Blok}, {Kennicutt}, {Schuster}, {Kramer}, {Wiesemeyer}, \&
  {Roussel}}]{Leroy09}
{Leroy}, A.~K., {Walter}, F., {Bigiel}, F., {et~al.} 2009, \aj, 137, 4670

\bibitem[{{Lu} {et~al.}(2017){Lu}, {Zhang}, {Kauffmann}, {Pillai}, {Longmore},
  {Kruijssen}, {Battersby}, {Liu}, {Ginsburg}, {Mills}, {Zhang}, \&
  {Gu}}]{Lu17}
{Lu}, X., {Zhang}, Q., {Kauffmann}, J., {et~al.} 2017, \apj, 839, 1

\bibitem[{{McMullin} {et~al.}(2007){McMullin}, {Waters}, {Schiebel}, {Young},
  \& {Golap}}]{McMullin07}
{McMullin}, J.~P., {Waters}, B., {Schiebel}, D., {Young}, W., \& {Golap}, K.
  2007, in Astronomical Society of the Pacific Conference Series, Vol. 376,
  Astronomical Data Analysis Software and Systems XVI, ed. R.~A. {Shaw},
  F.~{Hill}, \& D.~J. {Bell}, 127

\bibitem[{{Meier} {et~al.}(2014){Meier}, {Turner}, \& {Beck}}]{Meier14}
{Meier}, D.~S., {Turner}, J.~L., \& {Beck}, S.~C. 2014, \apj, 795, 107

\bibitem[{{Milam} {et~al.}(2005){Milam}, {Savage}, {Brewster}, {Ziurys}, \&
  {Wyckoff}}]{Milam05}
{Milam}, S.~N., {Savage}, C., {Brewster}, M.~A., {Ziurys}, L.~M., \& {Wyckoff},
  S. 2005, \apj, 634, 1126

\bibitem[{{Morokuma-Matsui} {et~al.}(2015){Morokuma-Matsui}, {Sorai},
  {Watanabe}, \& {Kuno}}]{Morokuma15}
{Morokuma-Matsui}, K., {Sorai}, K., {Watanabe}, Y., \& {Kuno}, N. 2015, \pasj,
  67, 2

\bibitem[{{Muller} {et~al.}(2014){Muller}, {Mizuno}, {Minamidani}, {Kawamura},
  {Rosie Chen}, {Indebetouw}, {Enokiya}, {Fukui}, {Gordon}, {Hayakawa},
  {Mizuno}, {Murai}, {Okuda}, {Onishi}, {Tachihara}, {Takekoshi}, {Yamamoto},
  \& {Yoshiike}}]{Muller14}
{Muller}, E., {Mizuno}, N., {Minamidani}, T., {et~al.} 2014, \pasj, 66, 4

\bibitem[{{Muraoka} {et~al.}(2012){Muraoka}, {Tosaki}, {Miura}, {Onodera},
  {Kuno}, {Nakanishi}, {Kaneko}, \& {Komugi}}]{Muraoka12}
{Muraoka}, K., {Tosaki}, T., {Miura}, R., {et~al.} 2012, \pasj, 64, 3

\bibitem[{{Muraoka} {et~al.}(2009){Muraoka}, {Kohno}, {Tosaki}, {Kuno},
  {Nakanishi}, {Handa}, {Sorai}, {Ishizuki}, \& {Okuda}}]{Muraoka09}
{Muraoka}, K., {Kohno}, K., {Tosaki}, T., {et~al.} 2009, \pasj, 61, 163

\bibitem[{{Muraoka} {et~al.}(2016){Muraoka}, {Sorai}, {Kuno}, {Nakai},
  {Nakanishi}, {Takeda}, {Yanagitani}, {Kaneko}, {Miyamoto}, {Kishida},
  {Hatakeyama}, {Umei}, {Tanaka}, {Tomiyasu}, {Saita}, {Ueno}, {Matsumoto},
  {Salak}, \& {Morokuma-Matsui}}]{Muraoka16}
{Muraoka}, K., {Sorai}, K., {Kuno}, N., {et~al.} 2016, \pasj, 68, 89

\bibitem[{{Murphy} {et~al.}(2015){Murphy}, {Dong}, {Leroy}, {Momjian},
  {Condon}, {Helou}, {Meier}, {Ott}, {Schinnerer}, \& {Turner}}]{Murphy15}
{Murphy}, E.~J., {Dong}, D., {Leroy}, A.~K., {et~al.} 2015, \apj, 813, 118

\bibitem[{{Nikiel-Wroczy{\'n}ski} {et~al.}(2013){Nikiel-Wroczy{\'n}ski},
  {Soida}, {Urbanik}, {We{\.z}gowiec}, {Beck}, {Bomans}, \&
  {Adebahr}}]{Nikiel13}
{Nikiel-Wroczy{\'n}ski}, B., {Soida}, M., {Urbanik}, M., {et~al.} 2013, \aap,
  553, A4

\bibitem[{{Paladino} {et~al.}(2009){Paladino}, {Murgia}, \&
  {Orr{\~a}}}]{Paladino09}
{Paladino}, R., {Murgia}, M., \& {Orr{\~a}}, E. 2009, \aap, 503, 747

\bibitem[{{Paladino} {et~al.}(2008){Paladino}, {Murgia}, {Tarchi},
  {Moscadelli}, \& {Comito}}]{Paladino08}
{Paladino}, R., {Murgia}, M., {Tarchi}, A., {Moscadelli}, L., \& {Comito}, C.
  2008, \aap, 485, 679

\bibitem[{{Peng} {et~al.}(1998){Peng}, {Ho}, {Filippenko}, \&
  {Sargent}}]{Peng98}
{Peng}, C.~Y., {Ho}, L.~C., {Filippenko}, A.~V., \& {Sargent}, W.~L.~W. 1998,
  in Bulletin of the American Astronomical Society, Vol.~30, American
  Astronomical Society Meeting Abstracts, 1253

\bibitem[{{Pety} {et~al.}(2013){Pety}, {Schinnerer}, {Leroy}, {Hughes},
  {Meidt}, {Colombo}, {Dumas}, {Garc{\'{\i}}a-Burillo}, {Schuster}, {Kramer},
  {Dobbs}, \& {Thompson}}]{Pety13}
{Pety}, J., {Schinnerer}, E., {Leroy}, A.~K., {et~al.} 2013, \apj, 779, 43

\bibitem[{{Pineda} {et~al.}(2008){Pineda}, {Caselli}, \& {Goodman}}]{Pineda08}
{Pineda}, J.~E., {Caselli}, P., \& {Goodman}, A.~A. 2008, \apj, 679, 481

\bibitem[{{Ptak} {et~al.}(2006){Ptak}, {Colbert}, {van der Marel}, {Roye},
  {Heckman}, \& {Towne}}]{Ptak06}
{Ptak}, A., {Colbert}, E., {van der Marel}, R.~P., {et~al.} 2006, \apjs, 166,
  154

\bibitem[{{Regan} {et~al.}(2002){Regan}, {Sheth}, {Teuben}, \&
  {Vogel}}]{Regan02}
{Regan}, M.~W., {Sheth}, K., {Teuben}, P.~J., \& {Vogel}, S.~N. 2002, \apj,
  574, 126

\bibitem[{{Regan} {et~al.}(2001){Regan}, {Thornley}, {Helfer}, {Sheth}, {Wong},
  {Vogel}, {Blitz}, \& {Bock}}]{Regan01}
{Regan}, M.~W., {Thornley}, M.~D., {Helfer}, T.~T., {et~al.} 2001, \apj, 561,
  218

\bibitem[{{Reuter} {et~al.}(1996){Reuter}, {Sievers}, {Pohl}, {Lesch}, \&
  {Wielebinski}}]{Reuter96}
{Reuter}, H.-P., {Sievers}, A.~W., {Pohl}, M., {Lesch}, H., \& {Wielebinski},
  R. 1996, \aap, 306, 721

\bibitem[{{Romano} {et~al.}(2017){Romano}, {Matteucci}, {Zhang},
  {Papadopoulos}, \& {Ivison}}]{Romano17}
{Romano}, D., {Matteucci}, F., {Zhang}, Z.-Y., {Papadopoulos}, P.~P., \&
  {Ivison}, R.~J. 2017, \mnras, 470, 401

\bibitem[{{Rots}(1978)}]{Rots78}
{Rots}, A.~H. 1978, \aj, 83, 219

\bibitem[{{Sage} {et~al.}(1991){Sage}, {Henkel}, \& {Mauersberger}}]{Sage91}
{Sage}, L.~J., {Henkel}, C., \& {Mauersberger}, R. 1991, \aap, 249, 31

\bibitem[{{Saintonge} {et~al.}(2011){Saintonge}, {Kauffmann}, {Kramer},
  {Tacconi}, {Buchbender}, {Catinella}, {Fabello}, {Graci{\'a}-Carpio}, {Wang},
  {Cortese}, {Fu}, {Genzel}, {Giovanelli}, {Guo}, {Haynes}, {Heckman},
  {Krumholz}, {Lemonias}, {Li}, {Moran}, {Rodriguez-Fernandez}, {Schiminovich},
  {Schuster}, \& {Sievers}}]{Saintonge11}
{Saintonge}, A., {Kauffmann}, G., {Kramer}, C., {et~al.} 2011, \mnras, 415, 32

\bibitem[{{Sakamoto} {et~al.}(1999){Sakamoto}, {Okumura}, {Ishizuki}, \&
  {Scoville}}]{Sakamoto99}
{Sakamoto}, K., {Okumura}, S.~K., {Ishizuki}, S., \& {Scoville}, N.~Z. 1999,
  \apjs, 124, 403

\bibitem[{{Scoville} \& {Solomon}(1974)}]{Scoville74}
{Scoville}, N.~Z., \& {Solomon}, P.~M. 1974, \apjl, 187, L67

\bibitem[{{Scoville} {et~al.}(1993){Scoville}, {Thakkar}, {Carlstrom}, \&
  {Sargent}}]{Scoville93}
{Scoville}, N.~Z., {Thakkar}, D., {Carlstrom}, J.~E., \& {Sargent}, A.~I. 1993,
  \apjl, 404, L59

\bibitem[{{Shirley}(2015)}]{Shirley15}
{Shirley}, Y.~L. 2015, \pasp, 127, 299

\bibitem[{{Sliwa} {et~al.}(2017){Sliwa}, {Wilson}, {Aalto}, \&
  {Privon}}]{Sliwa17}
{Sliwa}, K., {Wilson}, C.~D., {Aalto}, S., \& {Privon}, G.~C. 2017, \apjl, 840,
  L11

\bibitem[{{Smith} {et~al.}(1994){Smith}, {Harvey}, {Colome}, {Zhang},
  {Difrancesco}, \& {Pogge}}]{Smith94}
{Smith}, B.~J., {Harvey}, P.~M., {Colome}, C., {et~al.} 1994, \apj, 425, 91

\bibitem[{{Soida} {et~al.}(2001){Soida}, {Urbanik}, {Beck}, {Wielebinski}, \&
  {Balkowski}}]{Soida01}
{Soida}, M., {Urbanik}, M., {Beck}, R., {Wielebinski}, R., \& {Balkowski}, C.
  2001, \aap, 378, 40

\bibitem[{Sz\"{u}cs {et~al.}(2014)Sz\"{u}cs, Glover, \& Klessen}]{Szucs14}
Sz\"{u}cs, L., Glover, S. C.~O., \& Klessen, R.~S. 2014, Monthly Notices of the
  Royal Astronomical Society, 445, 4055.
\newblock \url{http://dx.doi.org/10.1093/mnras/stu2013}

\bibitem[{{Tacconi} {et~al.}(2013){Tacconi}, {Neri}, {Genzel}, {Combes},
  {Bolatto}, {Cooper}, {Wuyts}, {Bournaud}, {Burkert}, {Comerford}, {Cox},
  {Davis}, {F{\"o}rster Schreiber}, {Garc{\'{\i}}a-Burillo}, {Gracia-Carpio},
  {Lutz}, {Naab}, {Newman}, {Omont}, {Saintonge}, {Shapiro Griffin}, {Shapley},
  {Sternberg}, \& {Weiner}}]{Tacconi13}
{Tacconi}, L.~J., {Neri}, R., {Genzel}, R., {et~al.} 2013, \apj, 768, 74

\bibitem[{{Tacconi} {et~al.}(2018){Tacconi}, {Genzel}, {Saintonge}, {Combes},
  {Garc{\'{\i}}a-Burillo}, {Neri}, {Bolatto}, {Contini}, {F{\"o}rster
  Schreiber}, {Lilly}, {Lutz}, {Wuyts}, {Accurso}, {Boissier}, {Boone},
  {Bouch{\'e}}, {Bournaud}, {Burkert}, {Carollo}, {Cooper}, {Cox}, {Feruglio},
  {Freundlich}, {Herrera-Camus}, {Juneau}, {Lippa}, {Naab}, {Renzini},
  {Salome}, {Sternberg}, {Tadaki}, {{\"U}bler}, {Walter}, {Weiner}, \&
  {Weiss}}]{Tacconi17}
{Tacconi}, L.~J., {Genzel}, R., {Saintonge}, A., {et~al.} 2018, \apj, 853, 179

\bibitem[{{Tan} {et~al.}(2011){Tan}, {Gao}, {Zhang}, \& {Xia}}]{Tan11}
{Tan}, Q.-H., {Gao}, Y., {Zhang}, Z.-Y., \& {Xia}, X.-Y. 2011, Research in
  Astronomy and Astrophysics, 11, 787

\bibitem[{{Usero} {et~al.}(2015){Usero}, {Leroy}, {Walter}, {Schruba},
  {Garc{\'{\i}}a-Burillo}, {Sandstrom}, {Bigiel}, {Brinks}, {Kramer},
  {Rosolowsky}, {Schuster}, \& {de Blok}}]{Usero15}
{Usero}, A., {Leroy}, A.~K., {Walter}, F., {et~al.} 2015, \aj, 150, 115

\bibitem[{{van der Tak} {et~al.}(2007){van der Tak}, {Black}, {Sch{\"o}ier},
  {Jansen}, \& {van Dishoeck}}]{vanderTak07}
{van der Tak}, F.~F.~S., {Black}, J.~H., {Sch{\"o}ier}, F.~L., {Jansen}, D.~J.,
  \& {van Dishoeck}, E.~F. 2007, \aap, 468, 627

\bibitem[{{van Dishoeck} \& {Black}(1988)}]{Dishoeck88}
{van Dishoeck}, E.~F., \& {Black}, J.~H. 1988, \apj, 334, 771

\bibitem[{{Walter} {et~al.}(2008){Walter}, {Brinks}, {de Blok}, {Bigiel},
  {Kennicutt}, {Thornley}, \& {Leroy}}]{Walter08}
{Walter}, F., {Brinks}, E., {de Blok}, W.~J.~G., {et~al.} 2008, \aj, 136, 2563

\bibitem[{{Warren} {et~al.}(2010){Warren}, {Wilson}, {Israel}, {Serjeant},
  {Bendo}, {Brinks}, {Clements}, {Irwin}, {Knapen}, {Leech}, {Matthews},
  {M{\"u}hle}, {Mortimer}, {Petitpas}, {Sinukoff}, {Spekkens}, {Tan},
  {Tilanus}, {Usero}, {van der Werf}, {Vlahakis}, {Wiegert}, \&
  {Zhu}}]{Warren10}
{Warren}, B.~E., {Wilson}, C.~D., {Israel}, F.~P., {et~al.} 2010, \apj, 714,
  571

\bibitem[{{Watanabe} {et~al.}(2011){Watanabe}, {Sorai}, {Kuno}, \&
  {Habe}}]{Watanabe11}
{Watanabe}, Y., {Sorai}, K., {Kuno}, N., \& {Habe}, A. 2011, \mnras, 411, 1409

\bibitem[{{Welch} {et~al.}(1996){Welch}, {Thornton}, {Plambeck}, {Wright},
  {Lugten}, {Urry}, {Fleming}, {Hoffman}, {Hudson}, {Lum}, {Forster}, {Thatte},
  {Zhang}, {Zivanovic}, {Snyder}, {Crutcher}, {Lo}, {Wakker}, {Stupar},
  {Sault}, {Miao}, {Rao}, {Wan}, {Dickel}, {Blitz}, {Vogel}, {Mundy},
  {Erickson}, {Teuben}, {Morgan}, {Helfer}, {Looney}, {de Gues}, {Grossman},
  {Howe}, {Pound}, \& {Regan}}]{Welch96}
{Welch}, W.~J., {Thornton}, D.~D., {Plambeck}, R.~L., {et~al.} 1996, \pasp,
  108, 93

\bibitem[{{We{\.z}gowiec} {et~al.}(2012){We{\.z}gowiec}, {Soida}, \&
  {Bomans}}]{Soida12}
{We{\.z}gowiec}, M., {Soida}, M., \& {Bomans}, D.~J. 2012, \aap, 544, A113

\bibitem[{{Wilson}(1999)}]{Wilson99}
{Wilson}, T.~L. 1999, Reports on Progress in Physics, 62, 143

\bibitem[{{Wong} {et~al.}(2008){Wong}, {Ladd}, {Brisbin}, {Burton}, {Bains},
  {Cunningham}, {Lo}, {Jones}, {Thomas}, {Longmore}, {Vigan}, {Mookerjea},
  {Kramer}, {Fukui}, \& {Kawamura}}]{Wong08}
{Wong}, T., {Ladd}, E.~F., {Brisbin}, D., {et~al.} 2008, \mnras, 386, 1069

\bibitem[{{Yang} {et~al.}(2010){Yang}, {Stancil}, {Balakrishnan}, \&
  {Forrey}}]{Yang10}
{Yang}, B., {Stancil}, P.~C., {Balakrishnan}, N., \& {Forrey}, R.~C. 2010,
  \apj, 718, 1062

\bibitem[{{Yim} {et~al.}(2011){Yim}, {Wong}, {Howk}, \& {van der
  Hulst}}]{Yim11}
{Yim}, K., {Wong}, T., {Howk}, J.~C., \& {van der Hulst}, J.~M. 2011, \aj, 141,
  48

\bibitem[{{Yim} {et~al.}(2014){Yim}, {Wong}, {Xue}, {Rand}, {Rosolowsky}, {van
  der Hulst}, {Benjamin}, \& {Murphy}}]{Yim14}
{Yim}, K., {Wong}, T., {Xue}, R., {et~al.} 2014, \aj, 148, 127

\bibitem[{{Zhang} {et~al.}(1993){Zhang}, {Wright}, \& {Alexander}}]{Zhang93}
{Zhang}, X., {Wright}, M., \& {Alexander}, P. 1993, \apj, 418, 100

\bibitem[{{Zhang} {et~al.}(2014){Zhang}, {Henkel}, {Gao}, {G{\"u}sten},
  {Menten}, {Papadopoulos}, {Zhao}, {Ao}, \& {Kaminski}}]{Zhang14}
{Zhang}, Z.-Y., {Henkel}, C., {Gao}, Y., {et~al.} 2014, \aap, 568, A122

\end{thebibliography}

\end{document}